\documentclass[useAMS,usenatbib]{mnras}

 \usepackage{times}

\usepackage{graphicx}
\usepackage{supertabular}
\usepackage{subfig}
\usepackage{rotating}
\usepackage{lscape}
\usepackage{epsfig}
\usepackage{amssymb}
\usepackage{amsmath}
\usepackage{multicol}

\def\lesssim{\mathrel{\hbox{\rlap{\hbox{\lower4pt\hbox{$\sim$}}}\hbox{$<$}}}}
\def\gtrsim{\mathrel{\hbox{\rlap{\hbox{\lower4pt\hbox{$\sim$}}}\hbox{$>$}}}}

\def\mic{\,$\mu $m\,}

\def\apj{ApJ} 
\def\apjs{ApJSup} 
\def\mnras{MNRAS} 
\def\aaps{AAPS}
\def\pasp{PASP} 

\title[Multiband Algorithm for source Detection and
eXtraction.]{MADX - A simple technique for source detection and measurement using multi-band imaging from the Herschel-ATLAS survey.}
\author[S.J. Maddox]{S.J. Maddox$^{1,2}$
  \thanks{E-mail: maddoxs@cardiff.ac.uk} and L. Dunne$^{1,2}$\\
  $^{1}$School of Physics \&\ Astronomy, Cardiff University, Queen
  Buildings, The Parade, Cardiff, CF24 3AA, UK \\
  $^2$Institute for Astronomy, University of Edinbugh, Royal
  Observatory, Blackford Hill, Edinbugh EH9 3HJ, UK }
\begin{document}

\date{}

\pagerange{\pageref{firstpage}--\pageref{lastpage}} \pubyear{2019}

\maketitle

\label{firstpage}

\begin{abstract}
  We describe the method used to detect sources for the Herschel-ATLAS
  survey.  The method is to filter the individual bands using a matched
  filter, based on the point-spread function (PSF) and confusion
  noise, and then form the inverse variance weighted sum of the
  individual bands, including weights determined by a chosen spectral
  energy distribution. Peaks in this combined image are used to
  estimate the source positions. The fluxes for each source are
  estimated from the filtered single-band images, interpolated to the
  exact sub-pixel position.

  We test the method by creating simulated maps in three bands with
  PSFs, pixel sizes and Gaussian instrumental noise that match the
  250, 350 and 500\mic bands of Herschel-ATLAS.  We use our
  method to detect sources and compare the measured positions and
  fluxes to the input sources.  The multi-band approach allows
  reliable source detection a factor 1.2 to 3 lower in flux compared
  to single-band source detection, depending on the source
  colours. The false detection rate is reduced by a factor between 4
  and 10, and the variance of the source position errors is reduced by
  about a factor 1.5.  We also consider the effect of confusion noise
  and find that the appropriate matched filter gives a further
  improvement in completeness and noise over the standard PSF filter
  approach. Overall the two modifications give a factor of 1.5 to 3
  improvement in the depth of the recovered catalogues compared to a
  single-band PSF filter approach.

\end{abstract}


\section{Introduction}
\label{sec:intro} 

  There are many well-known algorithms to detect sources in imaging
data, from simple identification of connected pixels above a threshold
(\citealt{bertin}, \citealt{irwin}), through matched filtering
(\citealt{stetson}, \citealt{tegmark}, \citealt{herranz_matched}) to
wavelet techniques (\citealt{vielva}, \citealt{gonzalez-nuevo},
\citealt{grumitt}). These have generally been developed with single
pass-bands in mind, but recently the increase in availability of
multi-wavelength data has spurred the development of techniques that
make optimal use of several pass bands. This has been particularly
useful for sub-mm data (\citealt{naselsky}, \citealt{herranz},
\citealt{lanz}, \citealt{planck}). 

This paper describes the method that was used to detect sources
for the Herschel Astrophysical Terahertz Large Area Survey, hereafter
H-ATLAS (\citealt{eales}, \citealt{rigby}, \citealt{v16},
\citealt{m18}).  The H-ATLAS is based on observations in the 100, 160,
250, 350 and 500\mic bands of the {\it Herschel Space
  Observatory}\footnote{{\it Herschel} is an ESA space observatory
  with science instruments provided by European-led Principal
  Investigator consortia and with important participation from NASA}
(\citealt{Pilbratt}), which provide maps, covering $\sim$600 square degrees
of sky in the five bands. The unprecedented depth of the Herschel data
and the desire to have a blind far-infrared selected survey meant that we could
not rely on data from other surveys to identify sources; the source
detection had to be based on the Herschel maps alone. Also, the depth
of the Herschel data mean that the source density is high, and the
maps are significantly affected by source blending and confusion, so
standard methods do not perform well.

It is fairly straightforward to show that the optimal way to detect an
isolated point source in a map with a uniform background and simple
Gaussian noise, is to filter the data with the point-spread function
(PSF) and find the peak in the filtered map (e.g. \citealt{north},
\citealt{pratt}, \citealt{kay}, \citealt{stetson}).  The value of the
peak is equivalent to a least squares fit of the PSF to the data at
the position of the peak, and provides the minimum variance flux
estimate of the source. Our method is based on this matched filter
approach, but includes significant improvements: namely that the
matched filter includes the effect of confusion noise; the application
of the filter includes a locally defined noise-weighting; that several
bands can be combined in an optimal way to maximise the efficiency of
detecting sources; and that the fluxes are estimated sequentially to
reduce the effects of source blending.
Simply detecting images in each band individually and merging the
catalogues is not the optimal way to construct a combined catalogue,
and combining multi-wavelength data in an optimal way enhances the
source detection reliability and automatically produces a band-matched
catalogue. There has been extensive research in this area, considering
correlated noise between bands, variable source sizes and different
spectral behaviour, as reviewed by \cite{herranz_rev}. We developed
our method to find sources in the H-ATLAS survey, where data is
available in five bands, with different angular resolution, and each
with spatially varying noise. Note that the spatially non-uniform
noise distributions mean that it is not a good approximation to assume
simple Gaussian noise with known power spectra and cross-correlation
functions. This means that methods such as the matched multi-filter
approach of \cite{lanz} are not directly applicable to the
H-ATLAS data.

The next four sections in this paper describe the steps in the
detection and extraction process: in section~\ref{sec:back} we
estimate and subtract a non-uniform background; in
section~\ref{sec:filt} we filter the map for each waveband; in
section~\ref{sec:combine} we combine the wavebands and detect sources;
in section~\ref{sec:parameters} we estimate the source positions and
fluxes.  Then in section~\ref{sec:simulations} we describe simulations
which show the improvements of our method compared to single-band PSF
filtered catalogues.

\section{Background Estimation}
\label{sec:back} 

The first step in detecting sources is to estimate the background,
which may be spatially varying. A background may be instrumental or a
real astronomical signal which is a contamination to the point sources
we wish to extract. In the H-ATLAS, the background was largely local
`cirrus' from dust emission in our Galaxy. In general it is impossible
to differentiate between multiple confused sources and a smoothly
varying background (and/or foreground) component, but in either case,
it is necessary to remove the contribution of the background flux from
each individual source. So, we need to determine the local background
at all relevant positions in the map. We have done this by splitting
the map into blocks of pixels corresponding to $\sim 10\times$ the
full width at half maximum (FWHM) of the PSF, and constructing a
histogram of pixel values for each block. We then fit a Gaussian to
the peak of the histogram to find the modal value of the background,
and compare to the median value. If the peak is more than 1-$\sigma$
from the median, the fit is flagged as unreliable, and we use the
median instead. Near the edges of the map, there may be only a small
number of pixels contributing to a block. If there are less than 20
pixels in a block, the background is not estimated from the local
pixels, but is set to the final mean background from the whole
map. This ensures that the edges do not suffer from higher noise in
the background.

\begin{figure} %fig 1 
\includegraphics[scale=0.6]{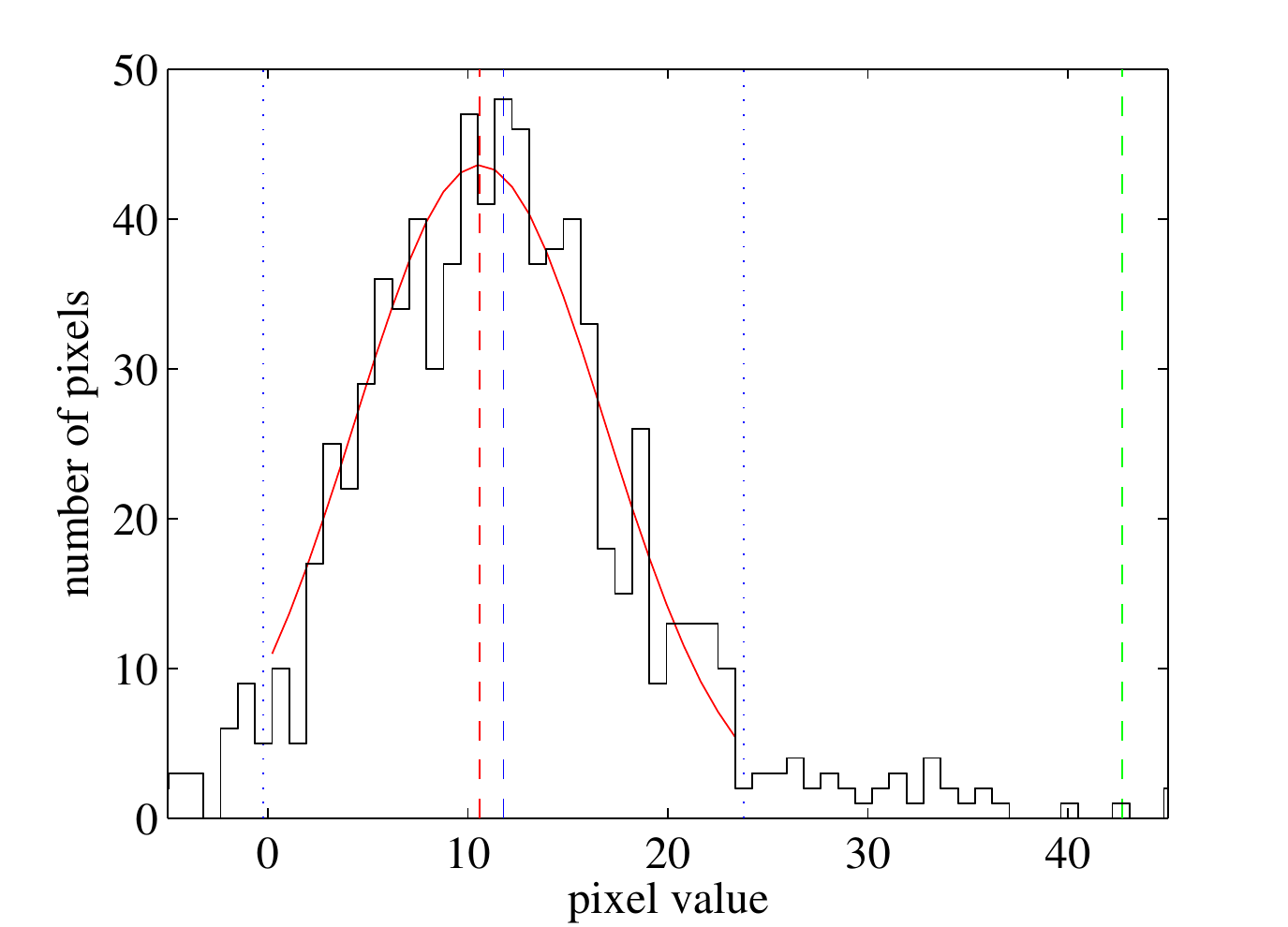}
\caption{\protect\label{fig:bhist_model} Simulated histogram of pixel
  values in a background block. The model has a true background of
  10mJy with 6mJy Gaussian noise and a single 1Jy source in the
  centre. The red line is the best fit Gaussian.  The fitted peak is
  is 10.3mJy (red dashed line), the median is 11.3mJy (blue dashed
  line), and the mean is 42mJy (green dashed line). The dotted blue
  lines are the $\pm 2 \sigma $ from the median.}
\end{figure}

This technique is valid only so long as the angular scale of a point
source is significantly smaller than the scale of background
variations. Since we have set the background blocks to be ten times
the FWHM of the PSF, this is a good approximation, and the fitted peak
of the histogram is very insensitive to bright sources in the
block. As a simple test we made a set of 1000 realisations of a model
with a background of 10mJy with Gaussian random noise with an rms of
6mJy, and put a single 1Jy Gaussian source in the middle. The
resulting histogram for a single realisation is shown in
Figure~\ref{fig:bhist_model}. The mean of the block is 42mJy, and so
would give an error of 32 mJy if it were used as the background
estimate. The median is more robust, leading to an error of 1mJy, and
the peak fit is biased by only 0.3 mJy.  It is worth noting that
background subtraction using simple filtering methods, such as the
Mexican Hat filter, are intrinsically linear, and so are approximately
equivalent to using the local mean value as the background
estimate. This means that they are significantly biased around bright
sources.

The background at each pixel is then estimated using a bi-cubic
interpolation between the coarse grid of backgrounds, and subtracted
from the data.  This approach to background estimation is similar to
the {\tt nebuliser}
algorithm\footnote{\url{http://casu.ast.cam.ac.uk/surveys-projects/software-release/background-filtering}},
developed by the Cambridge Astronomical Survey Unit. After the initial
analysis of the science-verification data for the H-ATLAS
survey (\citealt{rigby}), we used {\tt nebuliser} to perform the
background subtraction rather than the inbuilt background subtraction
(\citealt{v16}, \citealt{m18}). This choice was largely based on the
much faster run-time of {\tt nebuliser} compared to the MADX version.
The catalogues described in his paper use the built-in MADX background
subtraction. 

\section{Filtering}
\label{sec:filt} 

Typically image data is sampled finely enough that point sources have
2 or 3 pixels within the FWHM in each direction. This means that the
flux from a source is spread over many pixels, and the optimal
estimate of the source flux is given by a weighted sum over pixels.
For an isolated source on a uniform background with uniform Gaussian
errors, the minimum variance estimate of the source flux is given by
the sum of the data weighted by the point spread function (PSF) at the
true position of the source. Cross-correlating the data by the PSF
gives the PSF-weighted sum at the position of each pixel, and choosing
the peak in the filtered map gives the minimum variance estimate of
the source position and source flux (see e.g. \citealt{stetson}).

If the pixel uncertainties vary spatially, the optimal weighting must
also include the inverse of the estimated variance, as derived by
\cite{serjeant}.  If the power spectrum of the noise is not flat, the
optimal filter is different from the PSF. In
particular, when the source density is high, confusion noise is
important, and the optimal filter is narrower than the PSF. The
optimal filter, $Q$, can be estimated using a matched filter approach
that includes confusion noise (\citealt{chapin}). In this case, the
noise-weighted filtered map, $F$, is given by
\begin{equation}
F = \frac{(DW)\otimes Q}{W\otimes PQ} ,
\label{eqn:filt}
\end{equation}
where $D$ is the background subtracted data in each pixel, the weight
$W=1/\mathrm{var}(D)$ is the inverse of the variance of each pixel,
$P$ is the PSF, and $\otimes$ represents the cross-correlation operator.
Assuming that the instrumental noise on each pixel in the unfiltered
map is uncorrelated, the variance of each pixel in the filtered map is
given by
\begin{equation}
V = \frac{ W \otimes Q^2 }{(W \otimes PQ)^2},
\label{eqn:fvar}
\end{equation}
This is the generalisation of the PSF-filtering approach derived by
\cite{serjeant}; setting $Q=P$ yields exactly the Serjeant et
al. results. The noise-weighting in this step is particularly
important for the H-ATLAS data, since the noise varies dramatically on
small angular scales depending on the number of detector passes a
particular sky position has (\citealt{m18}).

The filtered map gives the minimum variance estimate of the flux that
a source would have at any given pixel in the map. Standard FFT
routines allow easy calculation of the filtered maps at integer pixel
positions, but in practice, sources are not centred on pixels. In
order to find the best flux estimates, we need to allow for sub-pixel
positioning. Without this, the fluxes will be significantly
underestimated, particularly when a source lies at the edge of a
pixel. Our approach to solving this problem is discussed in Section~\ref{sec:sub-pixel}.

\section{Combining Wavebands}
\label{sec:combine}

The filtered map in a single band provides an estimate of source flux
and uncertainty at any position, and this approach can be extended to
include any other wavebands that are available. If we know the
observed spectral energy distribution (SED), of a source,
$S(\lambda)$, then the flux in a band with response $R(\lambda)$, is
given by
\begin{equation}
  F = \frac{\int S(\lambda) R(\lambda) d\lambda } {\int R(\lambda)
    d\lambda } ,
\end{equation}
where we assume the detector measures total energy, as in a bolometer,
not the total count of photons as in a CCD. 
We define the normalised SED as $S_0(\lambda)$ where
\begin{equation}
  S_0(\lambda)= \frac{S(\lambda)}{ \int S(\lambda) d\lambda} ,
\end{equation}
so the observed SED of the source is $A S_{0}(\lambda)$, where
$A=\int S(\lambda) d\lambda$.  Given a set of filter pass-bands, $R_k$
and the source SED, the true broad-band flux in each band is
$F_k = AF_{k0}$, where
\begin{equation}
F_{k0} = \frac{\int S_{0}(\lambda) R_k(\lambda) d\lambda } {\int( R_k(\lambda)
  d\lambda } .
\end{equation}

Since the value of $A$ does not depend on wavelength, the filtered map
in each band gives an independent estimate of $A$. In order to combine
the maps, we need to have the estimates at exactly the same
position. As discussed in Section~\ref{sec:sub-pixel}, it is
reasonable to use a bi-cubic interpolation to estimate the source flux
at non-integer pixel positions. If we interpolate the lower resolution
maps to the pixel centres of the highest resolution map, then we can
take the inverse variance weighted sum to obtain the minimum variance
estimate of $A$ at the pixel positions of the highest resolution map.
For waveband $k$ the estimate of $A$ at position $x$ is
$A_k = F_k(x)/F_{k0}$, and the variance is
$\sigma_{A,k}^2 = V_k(x)/F_{k0}^2$. Hence the overall minimum variance
estimate of $A_{\mathrm{tot}}$ is given by
\begin{equation}
  \label{eqn:Atot} 
  A_{\mathrm{tot}} = \frac{\sum_k{F_{k}\frac{F_{k0}}{V_k}} }
  {\sum_k{\frac{F_{k0}^2}{V_k}}}, 
\end{equation}
and the uncertainty on $A_\mathrm{tot}$ is given by
\begin{equation}
  \label{eqn:sigmatot} 
\sigma_A^2 = \frac{1}{\sum_k{\frac{F_{k0}^2}{V_k}}}.
\end{equation}
So, the significance of a source detection at any position is
$A_\mathrm{tot}/\sigma_A $. This is a very simple derivation of the
standard result first presented by \cite{naselsky}. As for the case of
a filtered map in a single band, we estimate the most likely position
of the source as the position of the peak in the combined significance
map.

Note that these formulae include a factor $F_{k0}^2$ as part of the
weight given to the waveband $k$, so that the true SED acts as a
weighting term for each band as well as the inverse variance
factor. This makes intuitive sense: if a source's flux is expected to
peak in a particular band, we should give that band the most weight in
determining the position of the source; if a source has a flat
spectrum, so that the flux is equal in all bands, then all bands are
given equal weight. In general we do not actually know the true SED of
any particular source, and clearly don't know the SED of sources that
we have not yet detected. However, we can maximise the detection rate
of a particular type of source by choosing an SED prior to match.

This derivation considers an isolated source, but we can filter and
combine the full area of available data to produce a global
significance map, and find all the peaks to consider as potential
sources. For H-ATLAS, we kept those that are more than $2.5\sigma$ as
potential sources.  In principle we could retain all peaks, but
rejecting the low-significance peaks gives a large saving in
computing time, while not losing any significant detections.

\section{Source parameters}
\label{sec:parameters}
\subsection{Estimating positions and fluxes}

To estimate the position of each source, we perform a variance
weighted least-squares fit of a Gaussian to the $5\times 5$ pixels
around each peak. The position of the peak is allowed to vary freely,
and is not constrained to be at integer pixel positions.  We fit only
to pixels near the peak to minimise the effects of confusion from
other nearby sources. Since the individual maps have been filtered,
the peak pixels already include data from the surrounding raw pixels,
combined in an optimal way; the peak fitting is solely to find the
position at the sub-pixel level.

In order to estimate the flux in each band, we use the individual
filtered maps, and interpolate to find the value of each map at the
position of the peak in the combined map. For an isolated source, this
will provide the optimal flux estimates. However, if there are sources
that are close together, so that the PSFs significantly overlap, this
simple approach will `double-count' flux, because the wings of each
source add on to the peak of its neighbour. A simple way to avoid this
problem is to sort the sources in order of decreasing flux based on
their initial peak-pixel value, and then estimate the optimal fluxes
in sequence. After getting the optimal fluxes for a source, we
subtract the scaled filtered PSF from the maps before estimating the
fluxes for the next source.  This is similar in concept to the clean
algorithm (\citealt{hogbom}) but with just one pass. This process is
done separately for each band, so the `clean' works from the brightest
sources in each band. In principle, the procedure could be iterated to
a stable solution, but in practice the difficult cases are blends of
sources that require a more sophisticated de-blending technique to
improve the flux estimates. So iterating this simple clean algorithm
would provide a very small gain in reliability at a large
computational cost.

To provide the uncertainty for each flux measurement, we create a map
of the filtered noise, using equation~\ref{eqn:fvar} and perform the
same interpolation to estimate the flux variance in each band at each
source position.

\subsection{Sub-pixel subtleties} 
\label{sec:sub-pixel} 

The above analysis ignores the complication that our data typically
sample the sky with only 2 or 3 pixels across the FWHM of the PSF.
When the PSF is sampled into coarse pixels, the value of the peak
pixel is averaged over the whole area of the central pixel, and so is
suppressed relative to the peak of the true PSF. For a PSF that is
close to a Gaussian with 3 pixels across the FWHM, this suppression is
typically $\sim5$ per\,cent.  Since we use the pixelated PSF (the
Point Response Function - or PRF) when filtering the data, the
filtered data is boosted by the suppression factor, and so the
estimated flux for source that is centred in a pixel is unbiased. For
a source that is not centred in a pixel, the observed peak value is
suppressed compared to a pixel centred source.

\begin{figure} %2 
\includegraphics[width=0.5\textwidth,trim={0 10mm 0 0},clip]{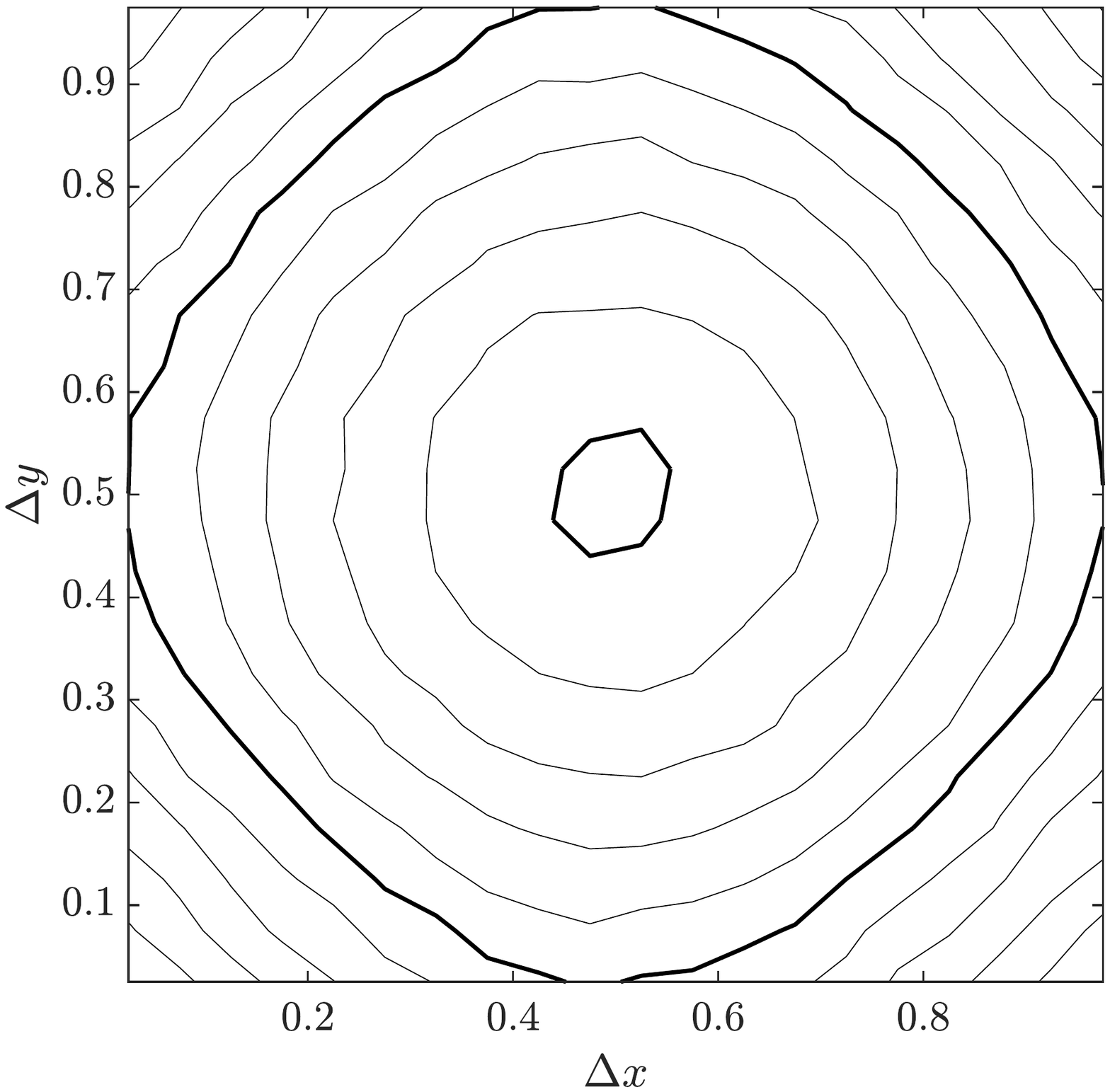}
\caption{Mean fractional flux errors as a function of the precise source
  position within a pixel. The contours are linearly space from 0 to
  $-0.9$ per\,cent. The central thicker contour corresponds to mean error of 0,
  and the outer   thicker contour is  $-0.5$ per\,cent. The maximum error is 
  $\sim -1$ per\,cent for a   source in the corner of 4 pixels. 
\label{fig:sub-pixel}
}
\end{figure}

In order to obtain the most accurate estimates of source flux for an
arbitrary position, we need to consider the true flux distribution
within the footprint of each pixel.  An obvious way to improve on the
flux estimated from individual pixel values is to interpolate between
them.  However, a bi-linear interpolation does not remove the bias, as
can be seen by considering a source that is exactly half way between
two pixels; each pixel will have the same value that is less than the
true peak value, so the interpolated value will also be biased low.  A
bi-cubic interpolation allows the interpolated value to be higher than
either individual pixel, and so gives a better flux estimate.

Since we estimate the source positions by fitting a Gaussian to the
central $5\times5$ pixels of each source, the positions are not
directly affected by the pixelization; for signal to noise ratio
greater than 20 we find that the source positions are accurate to
better than 1/12 of a pixel (see Figure~\ref{fig:positions}).  So, to
estimate the flux of the source, we use the filtered maps interpolated
to the best fit sub-pixel position.  If the source position lies at
the centre of a pixel, the interpolation returns that pixel value, and
this will be an unbiased flux estimate.  If the source position lies
at the boundary between two pixels, the pixel values are suppressed
relative to the peak of the pixel-centred PSF, but the bi-cubic
interpolation means that the estimated flux will be higher than the
pixel values, thus reducing the suppression. We tested this effect by
creating simulated data at higher resolution, averaging over the small
pixels to produce a low-resolution data, and then measuring the
recovered flux from the interpolated, filtered low-resolution data. We
find the interpolation reduces the suppression due to pixelization,
but does leave a slight underestimate of the actual peak.  The
fractional flux error as a function of position is shown in
Figure~\ref{fig:sub-pixel}. The error is zero at the centre of a
pixel, and smoothly increases towards the pixel edges. It is largest
for a source at the corner of 4 pixels when the flux is underestimated
by $\sim 1$ per\,cent. Although the simulations are specific to a
simple Gaussian PSF, this is a good approximation to the H-ATLAS
data. In fact, the PSF in most astronomical data can be approximated
by a Gaussian near the peak, and the pixel scales are typically chosen
to sample the FWHM at a similar spacing, and so similar improvements
are likely for other data.

\section{Tests of the methods} 
\label{sec:simulations}

\subsection{Simulations}

\begin{figure}% 3
\includegraphics[width=0.5\textwidth]{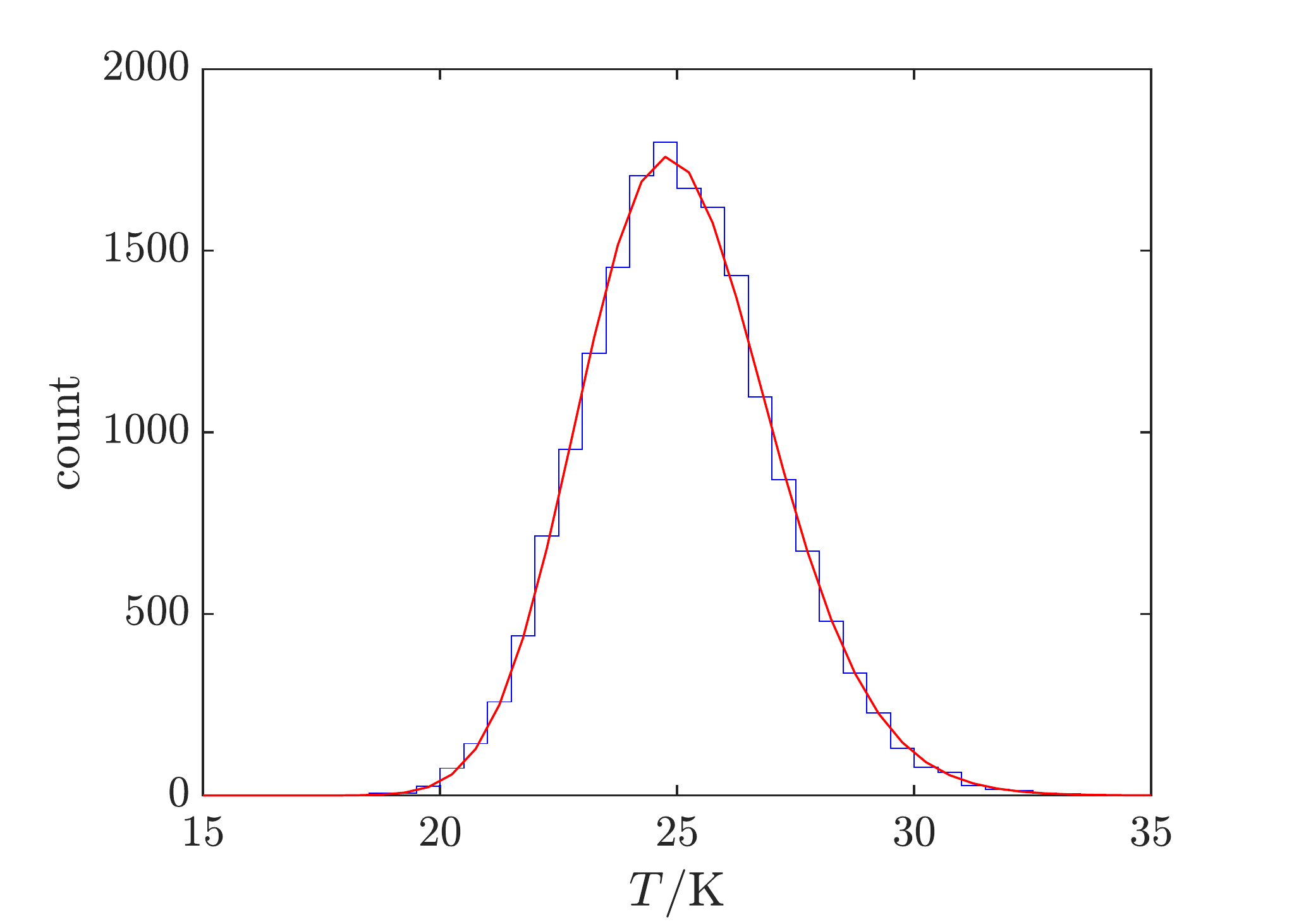} 
\caption{The temperature distribution of simulated sources. The red
  line shows the scaled and shifted log-normal probability distribution used to
  generate the temperatures. The  blue histogram shows the source
  counts for a single realisation of the simulations. 
  \label{fig:Tdist}}
\end{figure}

As a simple test of the source detection algorithm we generated maps
covering $3.4^\circ \times 13.6^\circ$, in three bands: 250, 350 and
500\mic.  This is equivalent to a single one of the three H-ATLAS
equatorial fields (\citealt{v16}). Sources are placed on a grid of
positions separated by 3 arcminutes on the sky with a small blank area
around the edges of the maps, leading to 17750 sources in the
maps. Each source is assigned a small random offsets from the exact
grid centre, so that the pixels sample the PSF with random offsets
from the pixel centres.  The PSF for each band is chosen to be a
Gaussian with full-width half maximum of 18, 24 and 36 arc seconds
respectively, roughly matching the Herschel beam in the three bands.
The PSF is over-sampled by a factor of 50, (corresponding to 0.12
arc-second pixels in the 250\mic band) and re-binned to the final
pixel sizes of 6, 8 and 12 arc seconds, chosen to match the H-ATLAS
maps.  Each source is given a 250\mic flux between 1mJy and 1Jy,
uniformly spaced in log flux. This is clearly not a good match to the
flux distribution of real sources, but is a simple way to provide good
statistics over the full flux range.  The 350\mic and 500\mic fluxes
for each source are then assigned so that the SED matches a modified
black body with $\beta$ chosen from a uniform random distribution
between 1 and 2, and temperature, $T$, randomly chosen from a
log-normal distribution centred on $T=25$K, and ranging from 20K to
35K, as shown in Figure~\ref{fig:Tdist}. This distribution roughly
matches the SEDs of low-redshift galaxies seen in the H-ATLAS survey
(\citealt{smith}).

\begin{figure}% 4
\includegraphics[width=0.5\textwidth]{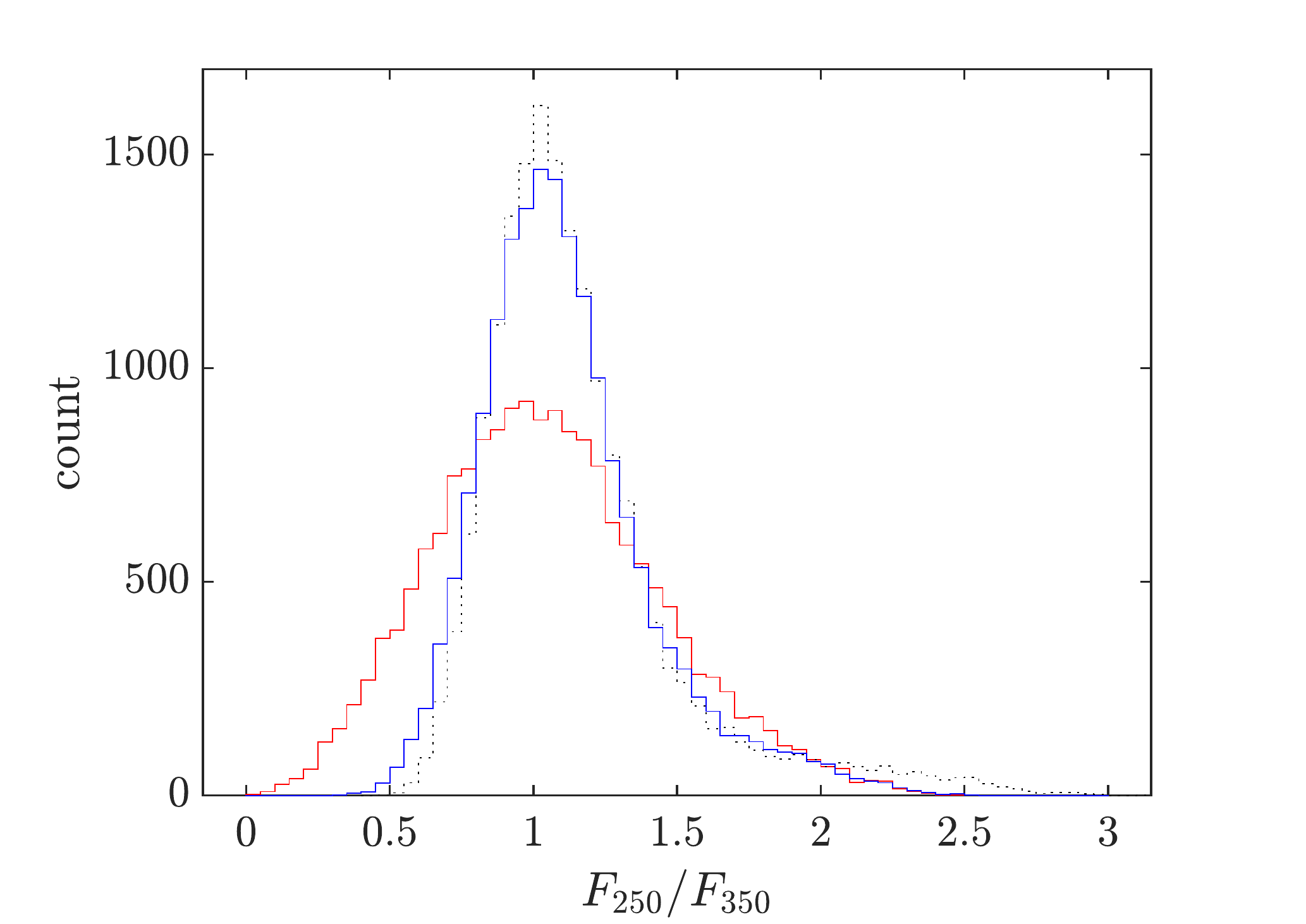} 
\caption{The $F_{250}/F_{350}$ colour  distribution of simulated sources
  compared to that observed in the H-ATLAS survey. The black dotted
  histogram shows the observed colour distribution. The  blue
  histogram shows the distribution for a particular realisation with
  matching colour distribution. The Red histogram shows a simulation
  with a extended high-z population . 
  \label{fig:col_hist}}
\end{figure}

Sources are included following a two-component redshift distribution with a
low-redshift population peaking at $z=0.3$, and a high redshift
population extending to $z\sim 2$ with a peak at $z=1.2$. This
reproduces the $F_{250}/F_{350}$ colour distribution observed in the
H-ATLAS survey, as shown in Figure~\ref{fig:col_hist}. Although these
simulations provide a reasonable match to the $F_{250}/F_{350}$ colour
of real data, they are not red enough in the $F_{500}/F_{350}$ colour
distribution. To investigate the impact for redder sources, we also
considered simulations with an extended tail of high redshift sources
which leads to an enhanced population of extremely red sources
(\citealt{highz}). Finding a best fit model which reproduces the
observed colour distributions is beyond the scope of this paper, but
the real H-ATLAS data will be somewhere between these two sets of 
simulations.


We also added a galactic background by taking the 100\mic and
temperature maps from \cite{SFD} and scaling the 100\mic emission
to the relevant wavelength using a modified black-body. The resolution
of these maps is several arcminutes, and so does not contain
small-scale structure in the cirrus background. The HATLAS data show
that in some patches of sky there is strong cirrus emission with
significant structure on sub-arc-minute scales, but in most areas, the
emission is relatively smooth. Our simulated background is a
reasonable approximation for most of the sky, but will be somewhat
easier to subtract than the areas where the true cirrus is
particularly strong and structured.

Finally we add Gaussian noise to the maps. The standard deviation is
set to roughly match the instrumental noise in the H-ATLAS survey
(\citealt{v16}). The values we use here are 9.3, 9.8 and 13.5\,mJy per
pixel in the 250, 350 and 500\mic bands respectively.  Note that the
sources are positioned on a grid, and so do not suffer from confusion
noise, meaning that the appropriate matched filter is the PSF. We consider
confusion noise and the modified matched filter in
Section~\ref{sec:mf}

We then run MADX on the simulated maps to detect the sources and
measure their position and fluxes. We used several different priors:
first using only the single bands for the detection: weights
1,0,0 for the 250\mic band; weights 0,1,0 for the 350\mic band;
weights 0,0,1 for the 500\mic band.  Second, we used equal weighting
for each band (weights 1,1,1), corresponding to a flat spectrum
source. By design, the sources are on a grid, and cannot overlap, so a
simple positional match allows us to associate the detected sources to
the corresponding input sources, and calculate the errors in position
and fluxes. We identify a recovered source with an input source if the
recovered position is within one pixel of the input position. For very
low signal-to-noise detections, the large standard deviation of the
positional errors means that some detections are not matched within
the one pixel radius. This has a small effect on the catalogue
completeness, but the dominant source of incompleteness is simply
noise on the flux estimates.

\subsection{Results}
\label{sec:results}
\begin{figure}
(a)\raisebox{-0.9\height}{\includegraphics[width=0.47\textwidth]{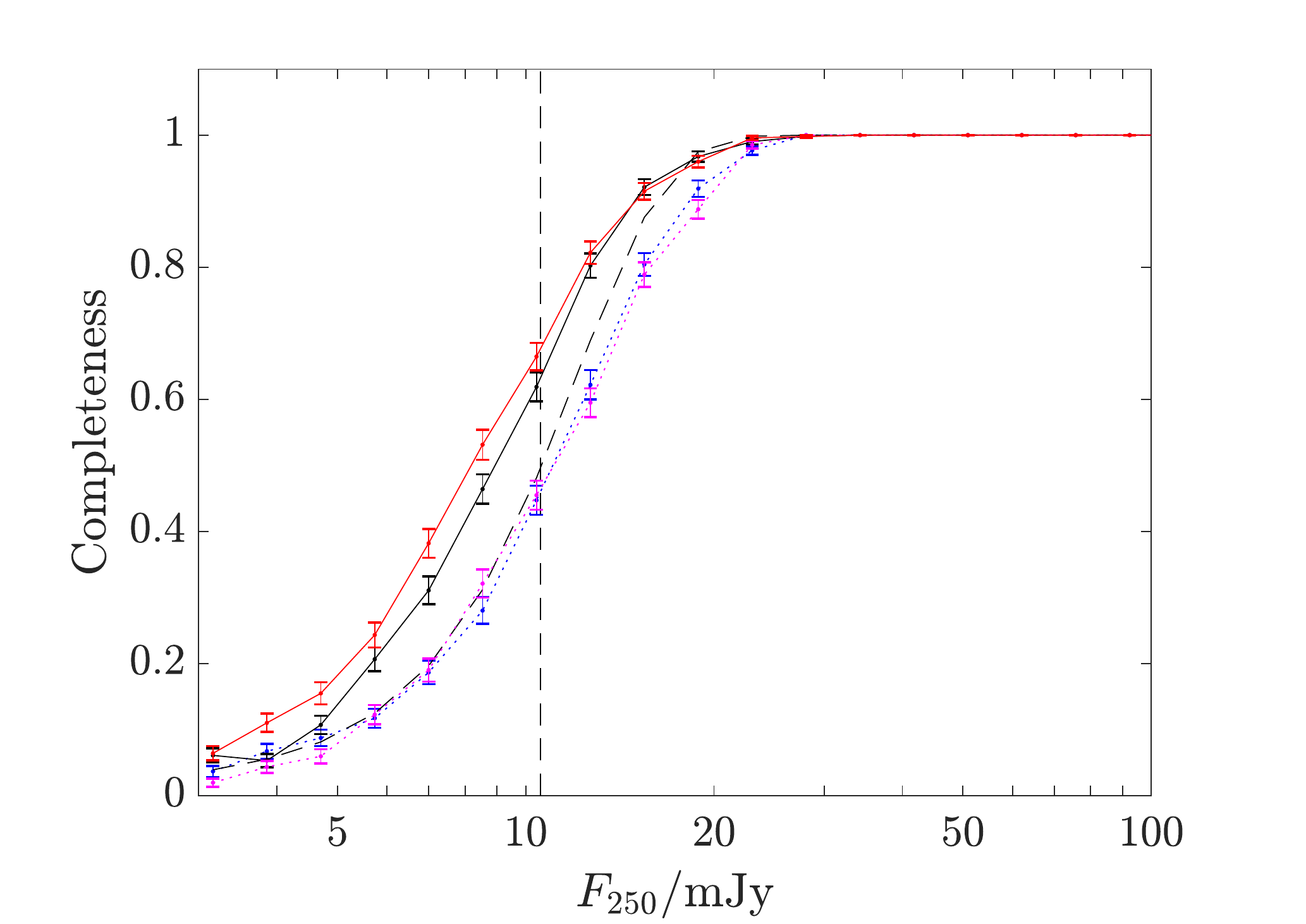}}\\
(b)\raisebox{-0.9\height}{\includegraphics[width=0.47\textwidth]{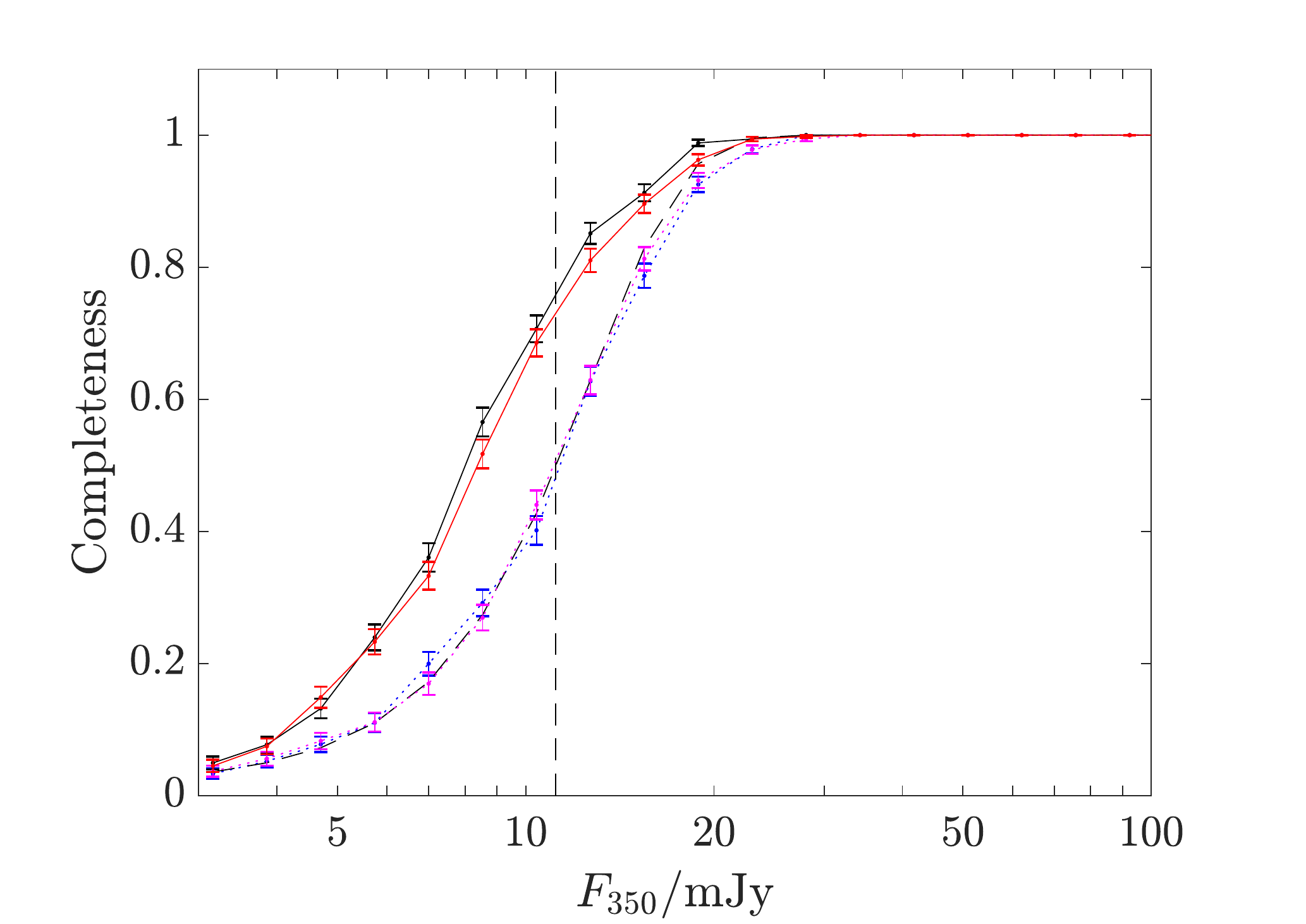}}\\
(c)\raisebox{-0.9\height}{\includegraphics[width=0.47\textwidth]{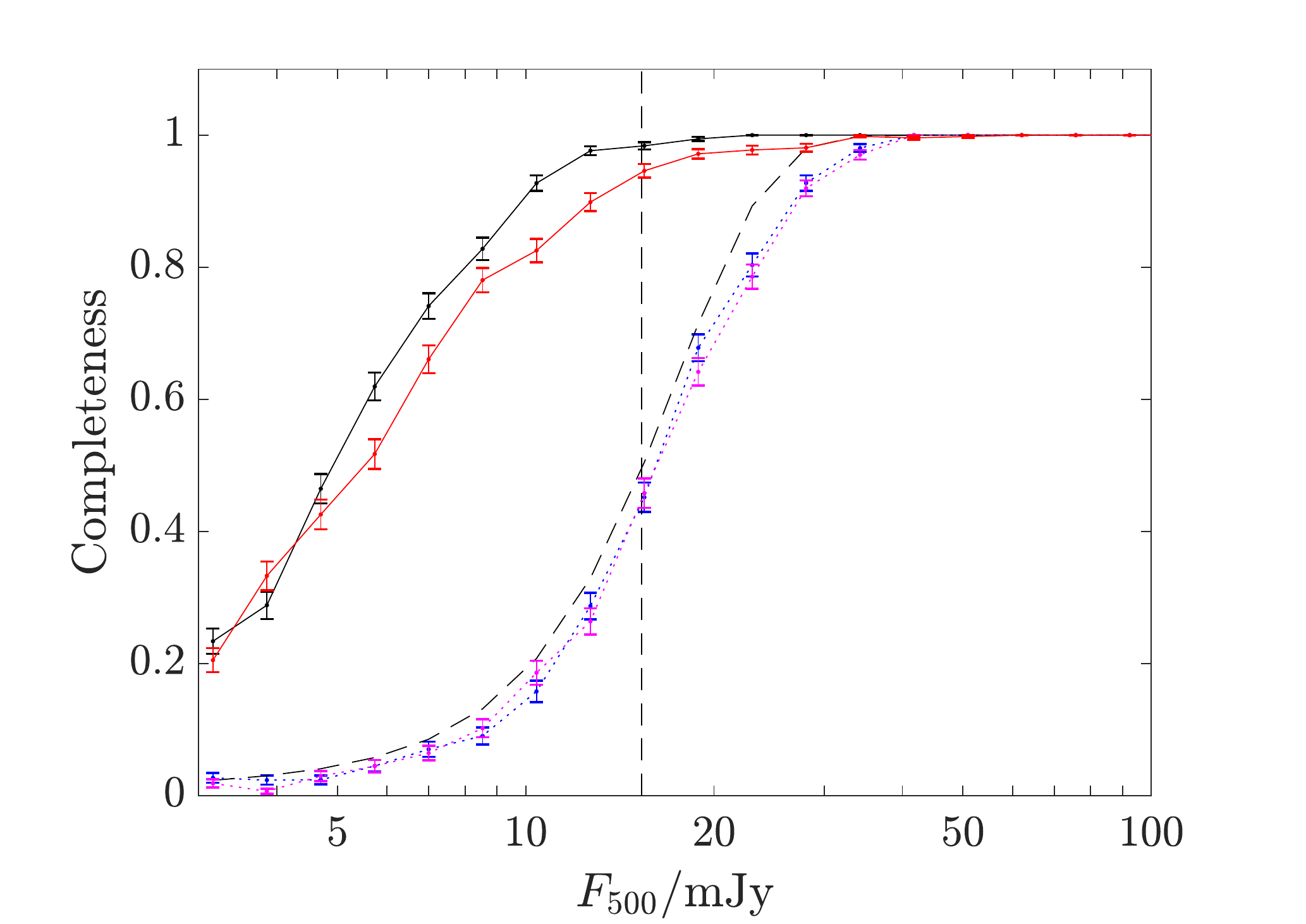}}
\caption{
Completeness of recovered source catalogues as a function of flux in
each band. The blue and black lines are for simulations which match
the H-ATLAS colour distribution; the red and magenta include a
high-redshift red population.  The dotted lines are for source
detection using only the relevant single band in each panel, and the
solid lines are for detection using the equal weighting of bands. The dashed curves show
the expected completeness from Gaussian errors and the vertical dashed
lines show the 2.5-$\sigma$ detection threshold. In the 250\mic band,
using the flat prior pushes the 50 per\,cent completeness limit about
a factor 1.25 deeper.  In the 350\mic and 500\mic bands the gains using the flat
prior are factors of 1.28 and 3 for the colour-matched simulations.
The effect of the red population is to slightly reduce the
completeness in the 350 and 500\mic bands. This is a small
effect as a percentage of the full catalogue, but represents a
significant improvement for the red population itself.
\label{fig:completeness} }
\end{figure}
\begin{figure}
(a)\raisebox{-0.9\height}{\includegraphics[width=0.47\textwidth]{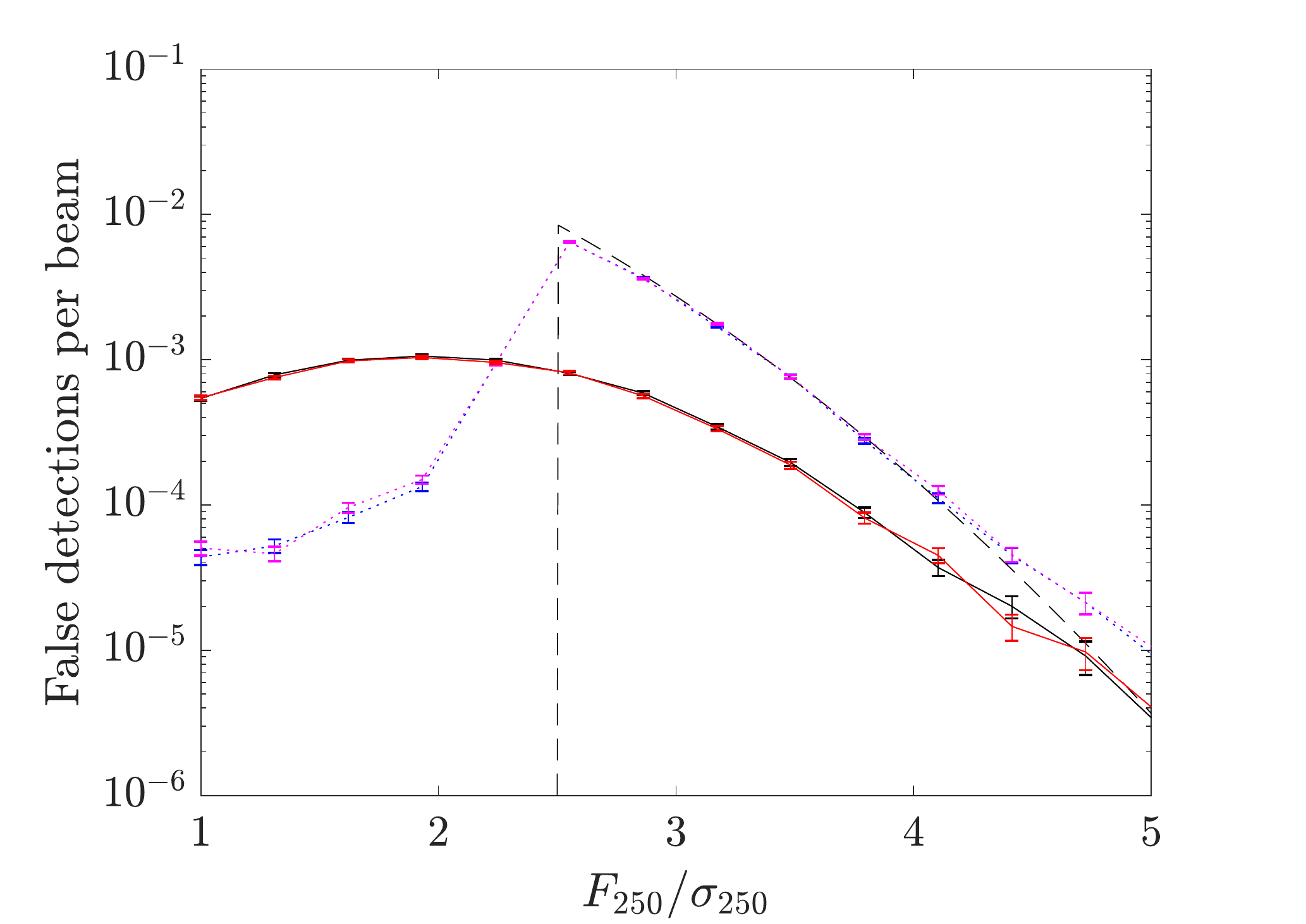}}\\
(b)\raisebox{-0.9\height}{\includegraphics[width=0.47\textwidth]{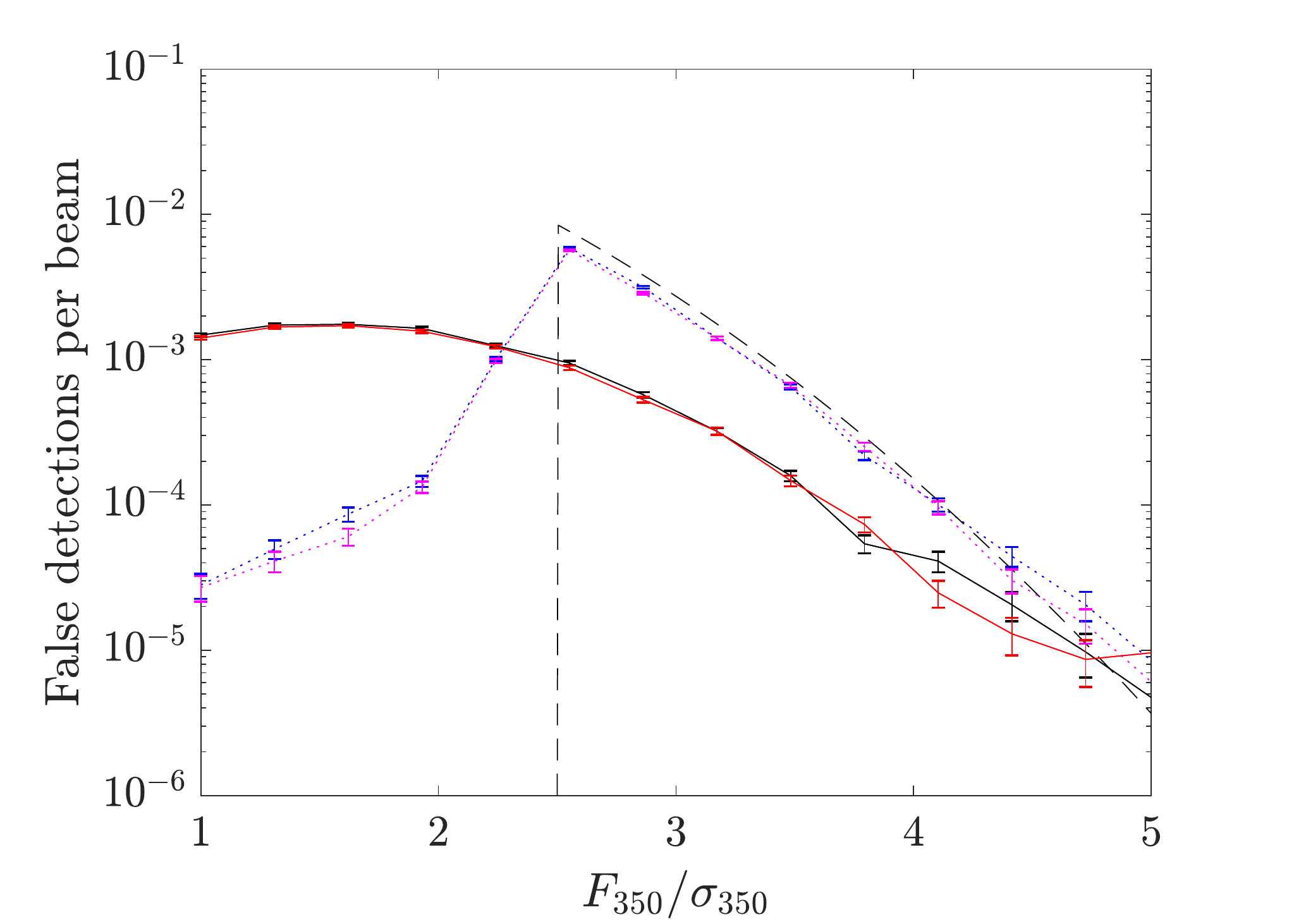}}\\
(c)\raisebox{-0.9\height}{\includegraphics[width=0.47\textwidth]{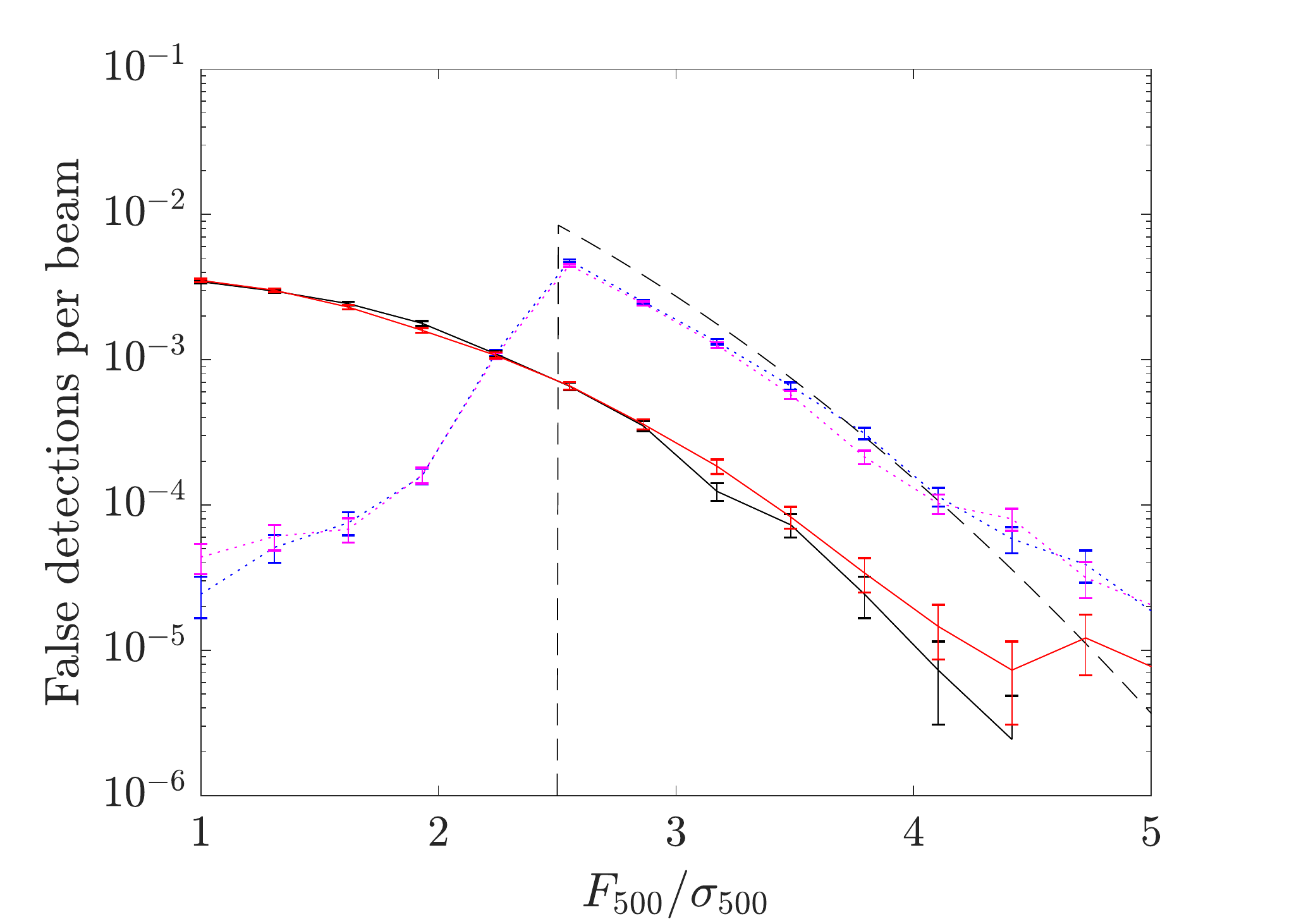}}
\caption{
 The number of false detections per beam as a function of signal to
noise. The blue and black lines are for simulations which match the
H-ATLAS colour distribution; the red and magenta include a
high-redshift red population.  The dotted lines are for source
detection using only the relevant single band and the solid 
lines show the equal weighting of bands.  The dashed lines 
show the expected rate of false detections from Gaussian errors and
the 2.5-$\sigma$ threshold applied during the MADX detection. For the
250\mic and 350\mic bands, the flat prior reduces the false detection rate by
about a factor 4 at the 4-$\sigma$ limit, and a factor 6 at
3-$\sigma$. In the 500\mic band the gain is roughly a factor 10 between
2 and 4-$\sigma$.
The inclusion of the extra red population does not significantly
change the false detection rates.  
\label{fig:false} }
\end{figure}

For each simulation, we measure the completeness by simply counting
the fraction of input sources that are detected as a function of
flux. Figure~\ref{fig:completeness} shows the completeness as a
function of flux in each band for both the colour-matched and
red-population simulations, and for catalogues using the single-band
priors and the flat priors.  The blue lines are for colour matched
simulations with catalogues using the single band prior and the black
lines use the flat prior.  It is clear that including information from
all three bands significantly improves the completeness of the
resulting catalogue. The flux limit at 50 per\,cent completeness in
the 250\mic and 350\mic band samples is a factor $\sim1.4$ deeper
using the multi-band approach compared to the single band method. The
gain for the 500\mic band is about a factor of 3, reflecting the very
large gain in signal to noise ratio by using the information from the
other bands to identify sources. The improvements are very similar for
the simulations with extra red sources.

The noise in the maps leads to peaks that are detected as a source,
but do not correspond to an input source.  The number of false
detections per beam area is shown as a function of signal to noise for
each band in Figure~\ref{fig:false}.  Using the flat-prior detection
reduces the number by a factor between 4 and 10 in all bands compared
to the single-band detection. Including extra red sources in the
simulations makes no significant difference to the rate of false
detections.  For the single band detections, the expected number of
noise peaks for Gaussian noise with a 2.5-$\sigma$ cut is shown by the
dashed line.  The observed number of false detections follows this
very well above the 2.5-$\sigma$ cut. There are a small number of
sources with fluxes below the 2.5-$\sigma$ limit, and this is because
cut is applied to the initial peak pixel value, while the final flux
plotted on $x$-axis uses the flux measured at the interpolated peak
position. The noise between the two flux estimates scatters some
sources below the initial cut.  In practise for the real H-ATLAS data
we apply a much higher limit (between 4 and 5$\sigma$) in the final
catalogue, so this effect is not visible.

\begin{figure}
(a)\raisebox{-0.9\height}{\includegraphics[width=0.47\textwidth]{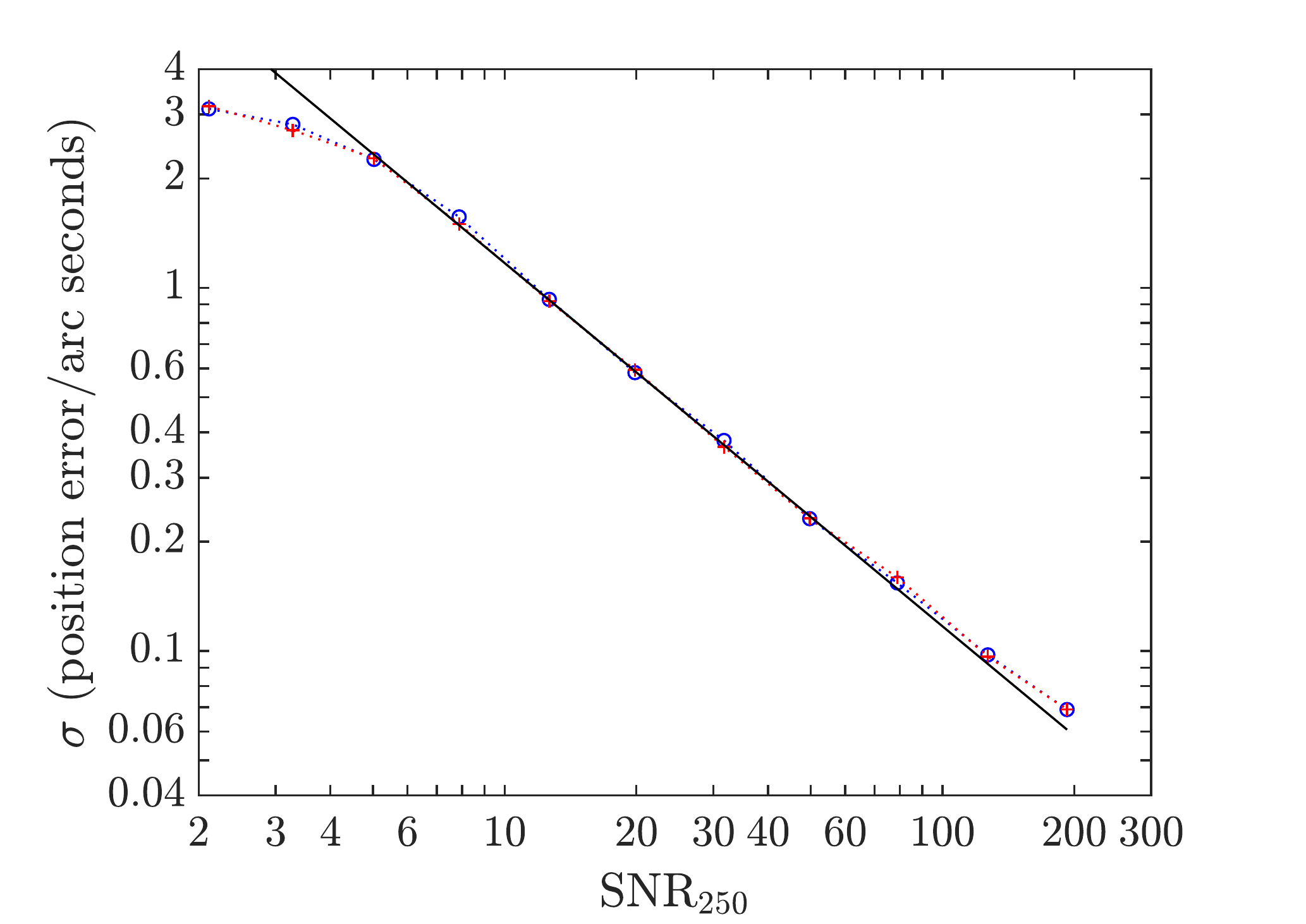}}\\ 
(b)\raisebox{-0.9\height}{\includegraphics[width=0.47\textwidth]{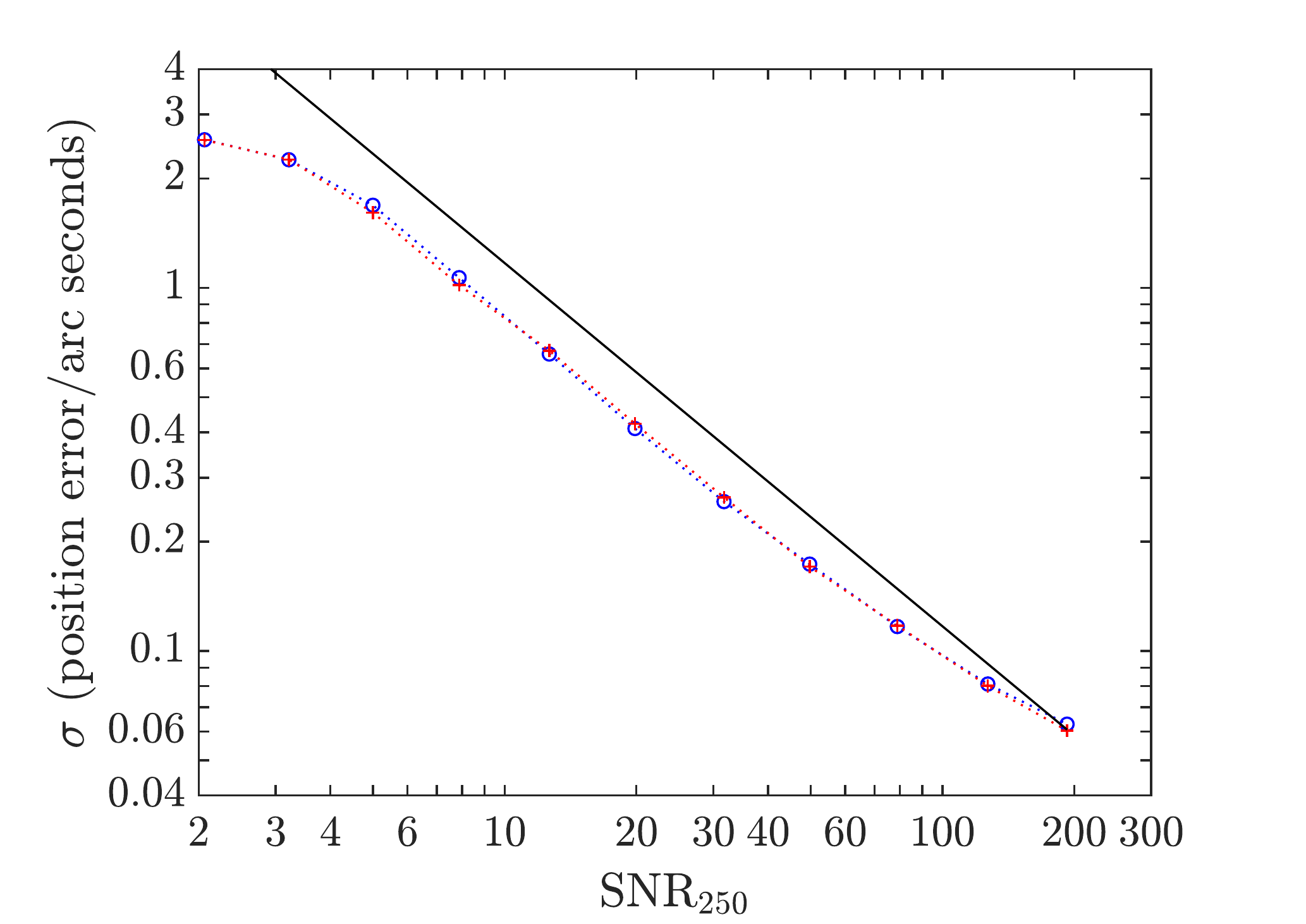}}
\caption{
\label{fig:positions} The measured positional errors of simulated
sources plotted as a function of signal-to-noise in the 250\mic
band. The standard deviation of the errors in RA are shown as blue
circles, and in Dec shown as red crosses.  Panel (a) shows the
measurements using only the 250\mic band to detect sources and measure
their positions. Panel (b) shows the measurements using the flat-prior
detection. The black line in both panels shows the variation expected
from the theoretical analysis of \protect\cite{ivison}. For
signal-to-noise ratio less than four the 1 pixel matching criterion
means that some true matches with large positional errors have not
been included in the matched sample, and this reduces the apparent
$\sigma$. A 2-d Gaussian with $\sigma=4$ truncated at 1 pixel radius
($=6''$) has a standard deviation of $2.7''$, as we see here.  }
\end{figure}

Next we compare the measured positions to the input positions. While
including data from lower resolution, or lower signal-to-noise bands
could increase the positional errors, extra information is provided by
summing data with the correct weighting, and the positional accuracy
is improved. This can be seen in Figure~\ref{fig:positions}, which
shows the rms position error in RA and Dec as a function of
signal-to-noise, defined as $A_{tot}/\sigma_A$ from
equations~\ref{eqn:Atot} and \ref{eqn:sigmatot}. Using only the
250\mic band to detect the sources leads to positional errors in good
agreement with the theoretical expectations from
\cite{ivison}. Including all bands significantly reduces the
positional errors, even though the other bands have poorer
resolution. At low signal-to-noise (SNR$\lesssim 4$) the apparent
positional error starts to flatten because only the sources within one
pixel are counted as matches, and the apparent standard deviation is
biased too low.

Finally we compare the measured and input fluxes for the sources, in
terms of both random and systematic errors. To assess the random
errors, we simply calculate the standard deviation of the between the
difference between the measured and input fluxes. Over the range of
interest, we find that this does not vary significantly as a function
of flux, and is 4.4, 4.6 and 6.4\,mJy for the 250\mic, 350\mic and
500\mic bands respectively. These are between 3 and 6 per\,cent higher
than the values of 4.2, 4.5 and 6.1\,mJy expected from the simple
application of equation~\ref{eqn:fvar} the estimate the flux
errors. This small discrepancy is likely to be caused by sub-pixel
positioning effects and residual background subtraction errors.  The
choice of prior makes no significant difference to the flux errors, as
these errors are dominated by the pixel-to-pixel flux errors on the
map for each band. Also the colour distribution used in the
simulations makes no significant difference to the errors.

\begin{figure}
(a)\raisebox{-0.9\height}{\includegraphics[width=0.47\textwidth]{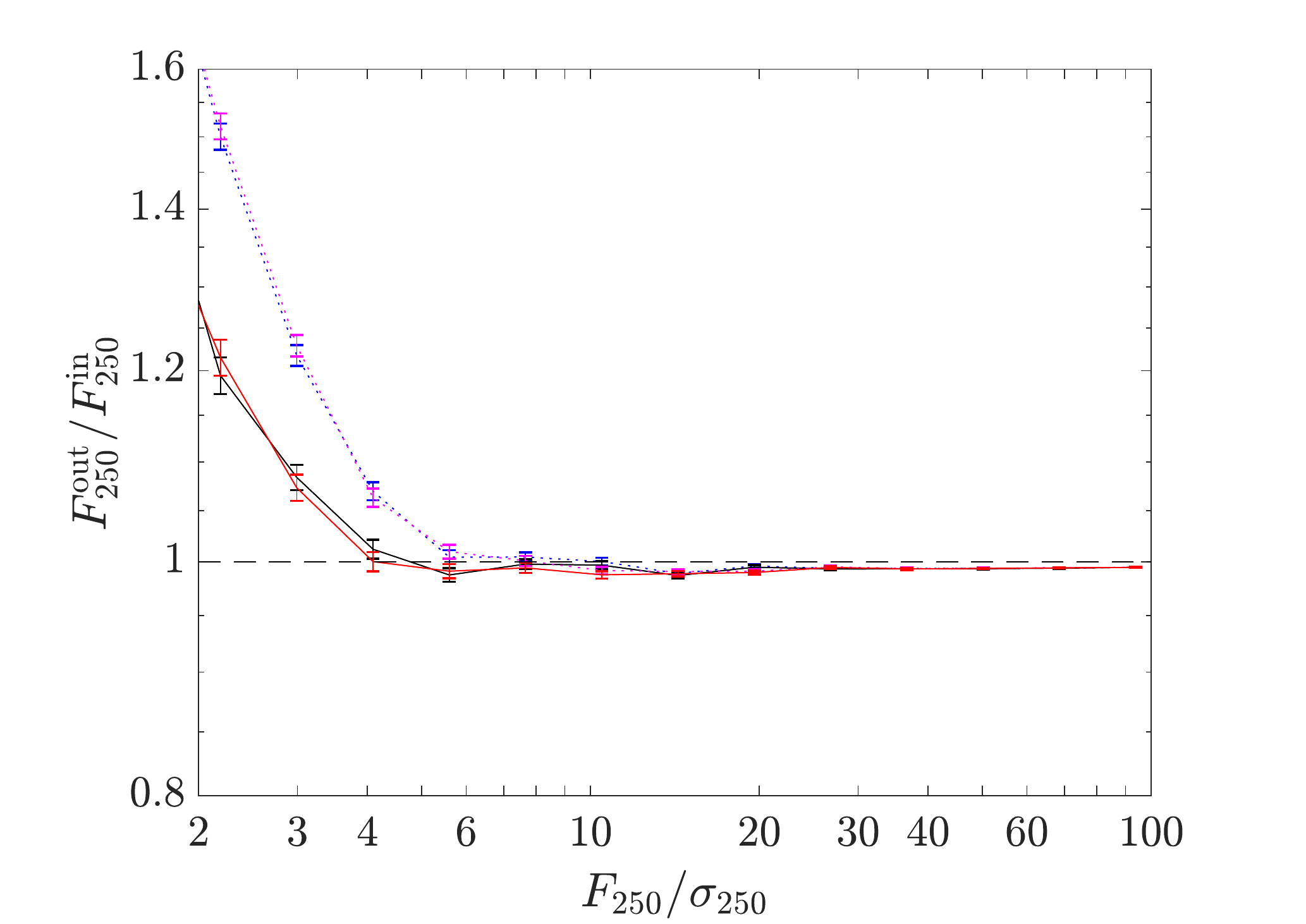}}
(b)\raisebox{-0.9\height}{\includegraphics[width=0.47\textwidth]{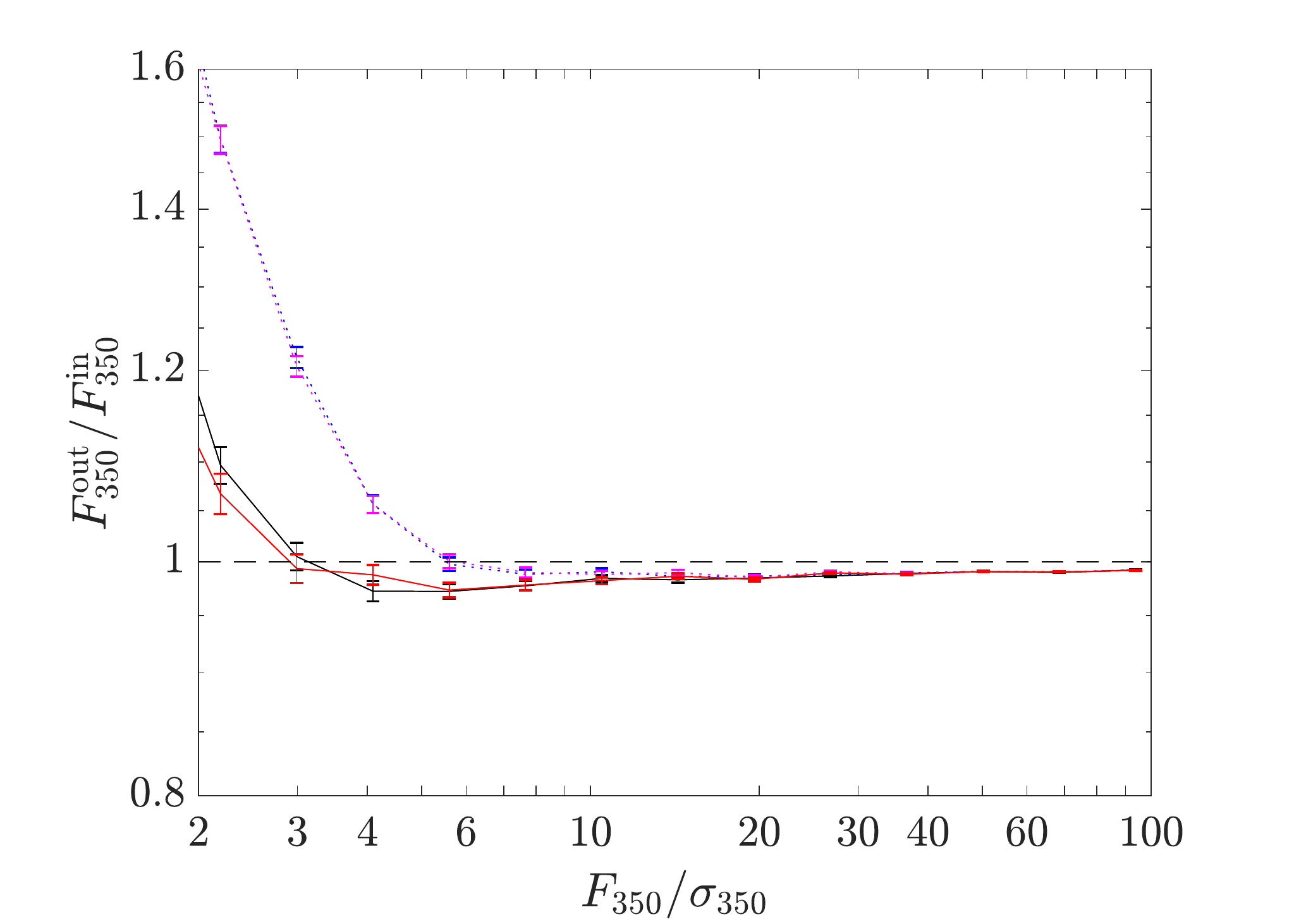}}
(c)\raisebox{-0.9\height}{\includegraphics[width=0.47\textwidth]{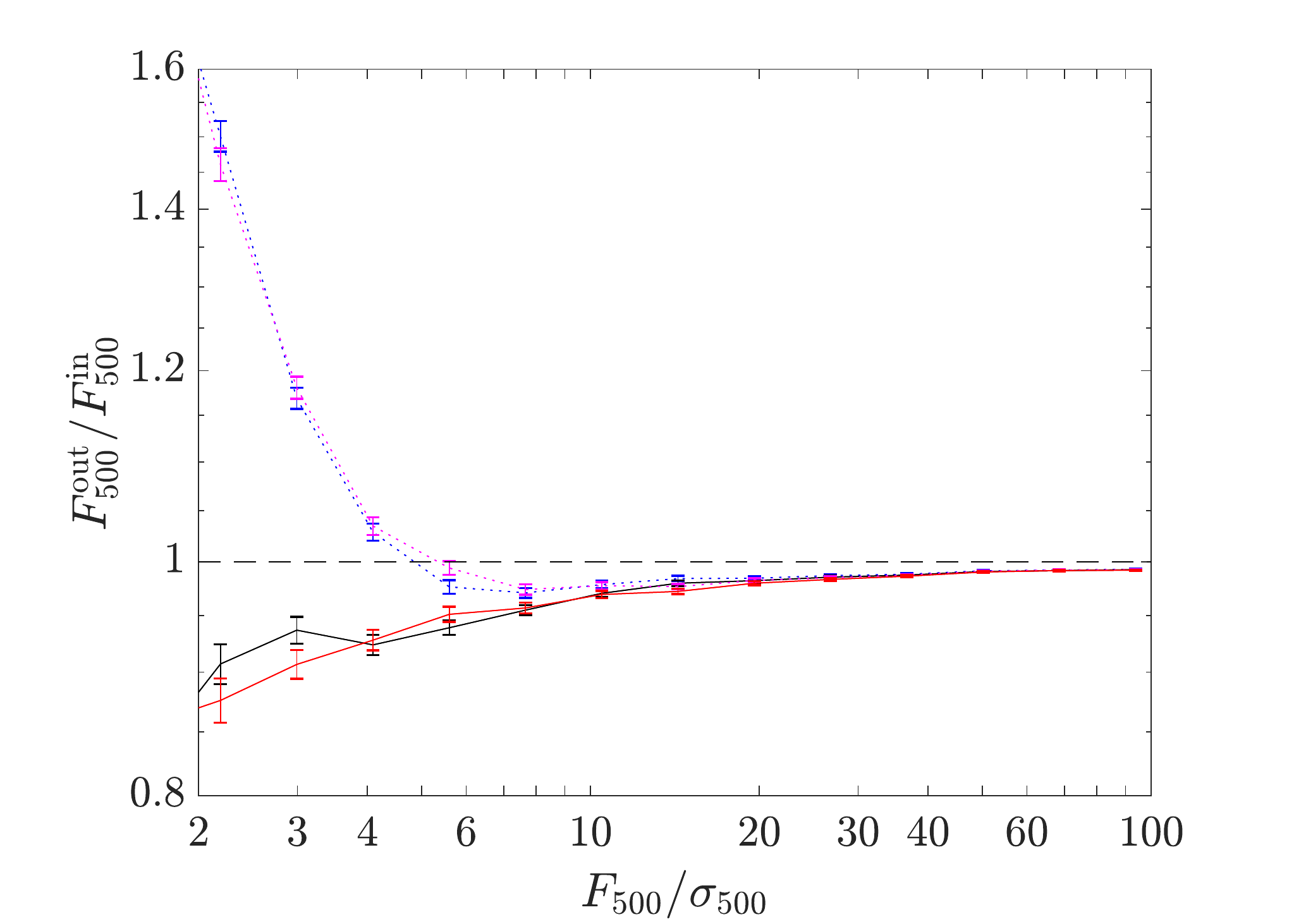}}
\caption{\label{fig:fluxes} The ratio of mean measured flux compared
  to the mean input flux as a function of input signal to noise for
  different source detection priors. The blue lines show the
  single-band priors, and the black lines use a flat prior. Panels
  (a),(b) and (c) shows the results for the 250\mic, 350\mic and
  500\mic fluxes respectively. Using the single-band source detection
  leads to significant flux boosting in the measured fluxes. For the
  250\mic and 350\mic bands, the flat prior reduces the boosting
  effect at fainter fluxes.  For the 500\mic band, the flat-prior
  fluxes are systematically underestimated at fainter fluxes.  The red
  and magenta lines show the results from simulations with extra red
  sources. The red sources do not introduce any significant changes. }
\end{figure}

Figure~\ref{fig:fluxes}(a) shows the mean ratio of measured to input
flux as a function of signal to noise. At high signal to
noise there is a small ($\sim$0.5 per\,cent) underestimate of flux due the
peak pixelization issues discussed in Section~\ref{sec:sub-pixel}.
For sources with fluxes near the detection limit, there is a
systematic bias to higher fluxes when using the single-bands to detect
sources (flux boosting). This is related to Eddington/Malmquist bias
when selecting sources to be above a signal-to-noise threshold: faint
sources with negative errors are not retained in the catalogue,
whereas those with positive errors are detected to a fainter level.
The precise form of the boosting depends on the distribution of true
source fluxes, as well as the measurement errors.  For our current
simulations we chose to distribute sources uniformly in log flux, and
so they do not match real source flux distributions, even though the
colours are realistic. Hence we cannot use them to estimate the
boosting as a function of flux for real data. \cite{v16} created
realistic simulations and used them to estimate both the completeness
and boosting correction factors that apply to the H-ATLAS data.

Using the flat prior combination to detect sources includes
information from all three bands, and so reduces the bias from noise
peaks in any single band, leading to significantly reduced flux
boosting. For the 500\mic band, the angular size of the PSF is larger,
and the signal-to-noise is significantly smaller than the other two
bands. This mean that it contributes only a small amount to the
detection signal and positional measurement. The positional errors
mean that the local peak is missed and the flux estimate is
systematically underestimated. This bias can be corrected using the
average values measured from simulations (\citealt{v16}).

The flux boosting and biases are not significantly affected by the
inclusion of extra red sources in the simulations.

\subsection{Confusion Noise}
\label{sec:mf}

\begin{figure}
(a)\raisebox{-0.9\height}{\includegraphics[width=0.47\textwidth]{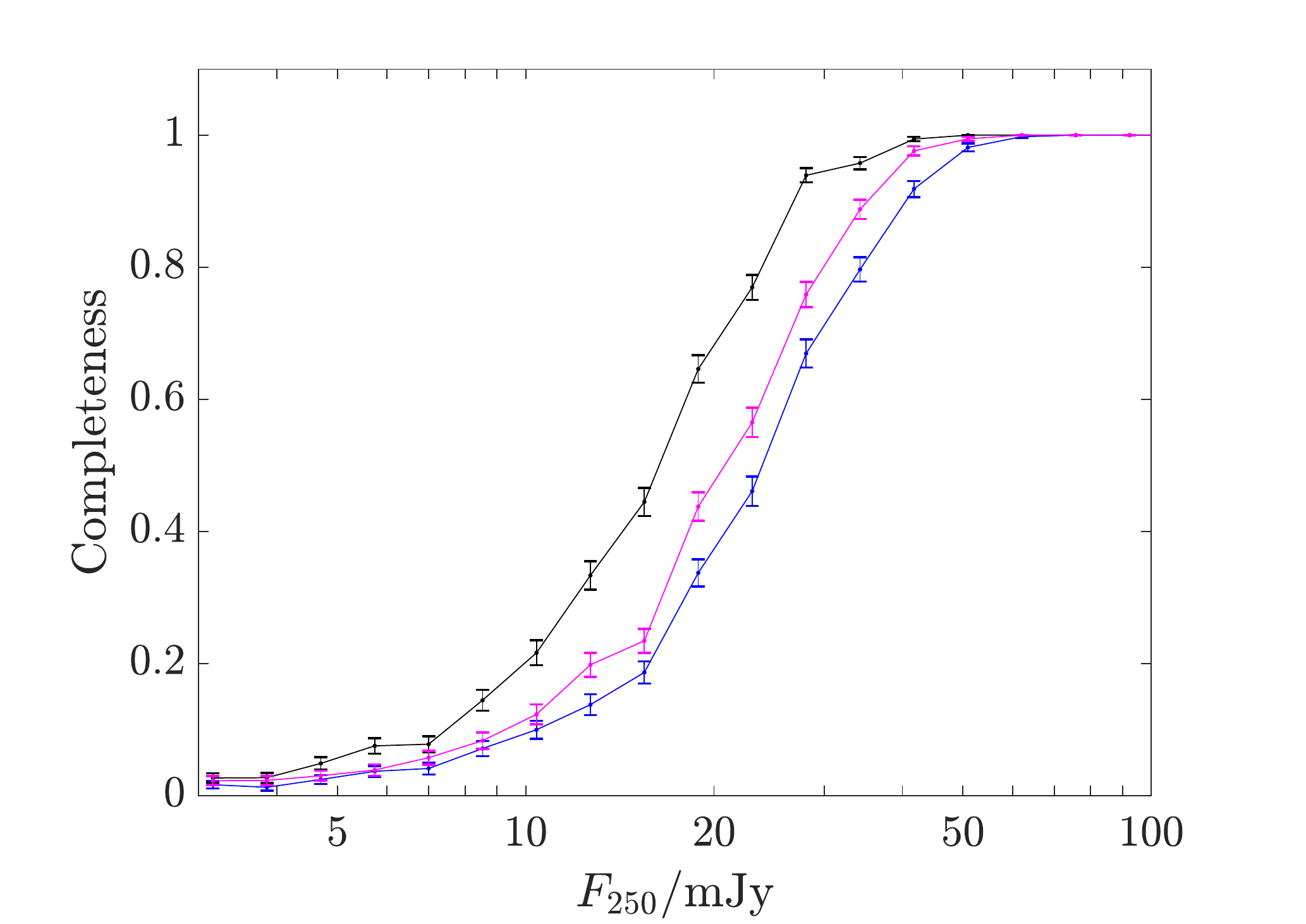}}
(b)\raisebox{-0.9\height}{\includegraphics[width=0.47\textwidth]{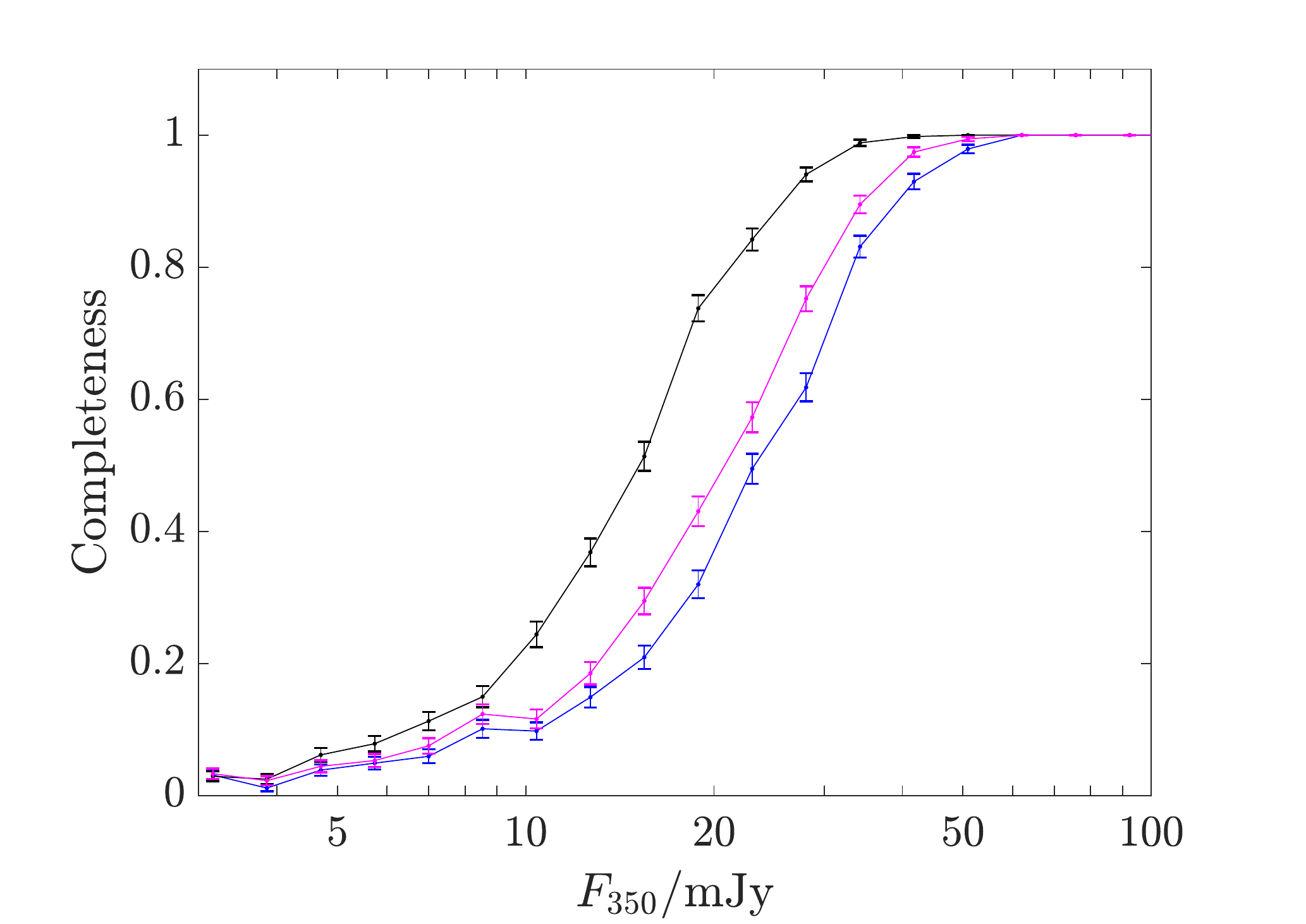}}
(c)\raisebox{-0.9\height}{\includegraphics[width=0.47\textwidth]{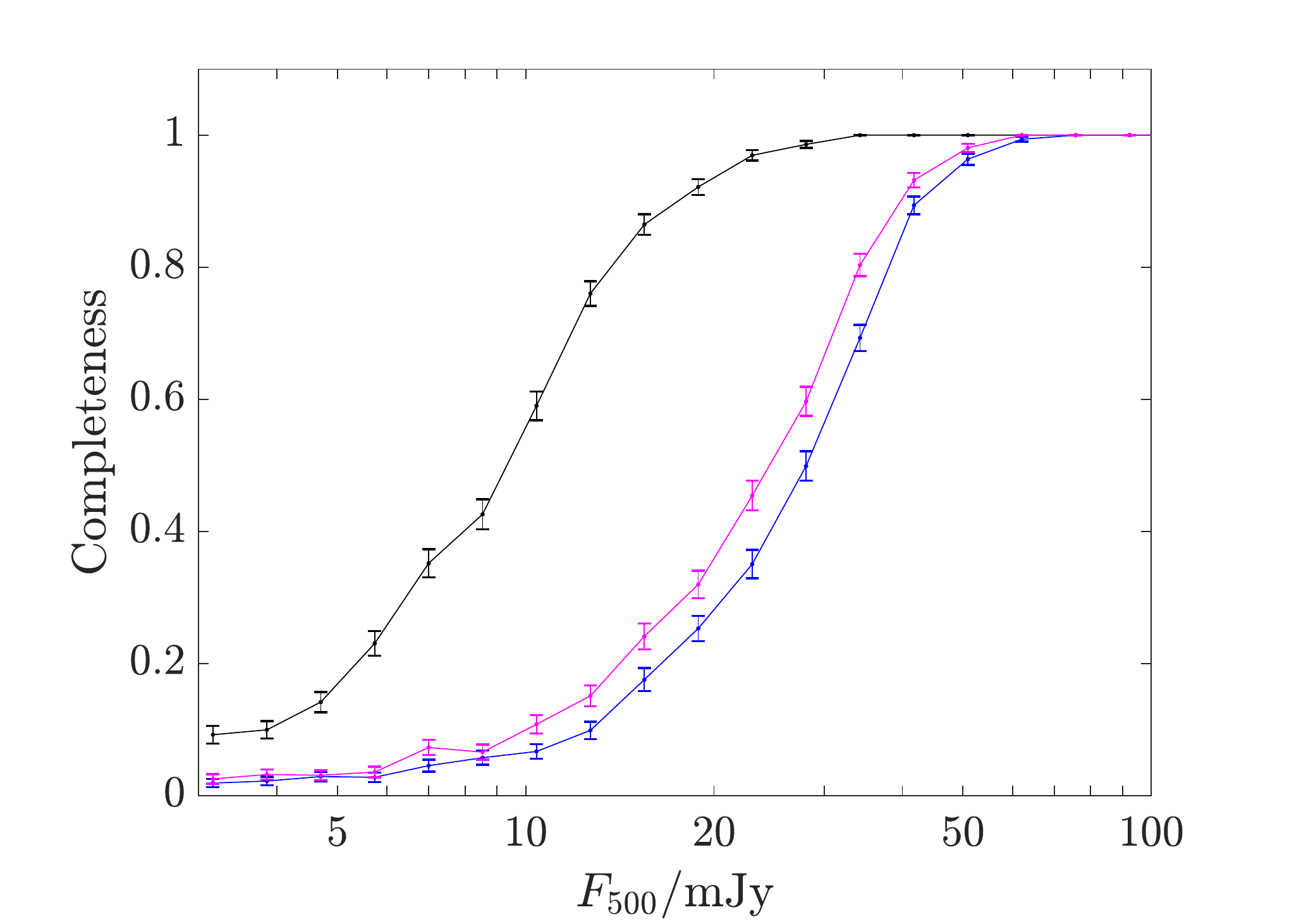}}
\caption{ Completeness of recovered source catalogue as a function of
flux in each band for simulations with confusion noise included. The
blue lines are for source detection using the appropriate single-band
prior and
PSF filtering; the magenta lines are for source detection using the
single-band priors, but with the matched filters; and the black lines use
the matched filters and equal weighting of bands. 
\label{fig:completeness_mf}
}
\end{figure}

The simulations described so far contain no confusion noise, and so
the appropriate optimal filter is simply the PSF. To test the
performance of the modified matched filter approach, we have added
confusion noise to each pixel in the simulations as an extra term
consisting of PSF-filtered Gaussian noise, with the variance as seen
in the H-ATLAS data (\citealt{v16}). The corresponding standard
deviation is $\sim$7\,mJy per pixel in all three bands. We use these
values of confusion noise to calculate the matched filters as
described in Appendix A of \cite{chapin}. We then re-ran the image
detection based on the single-band priors, and used both the PSF
filter and the matched filter. We also used the flat-prior detection
with the matched filters. The resulting completeness comparisons are
shown in Figure~\ref{fig:completeness_mf}. The matched filter
selection improves the completeness at a given signal to noise cut,
providing a catalogue $\sim$20 per\,cent deeper. As before, using the
flat prior detection also provides an improvement of a factor
$\sim1.3$ in flux at the same completeness level for the 250\mic and
350\mic bands, and a factor 2 for the 500\mic band. Overall the two
modifications give a catalogue that is a factor of 1.5 to 3 deeper in
flux at 50 per\,cent completeness.

The added confusion noise means that the flux errors are larger than
for the simulations with only Gaussian noise. Using the matched filter
reduces the errors by about 10 per\,cent compared to using the PSF. As an
aside, we note that using the matched filter for the Gaussian noise
simulations leads to a 14 per\,cent increase in flux errors compared to the
PSF filter. This is exactly as would be expected since the optimal
filter for data with Gaussian noise is the PSF, and the optimal filter
for data with confusion noise is the appropriate matched filter.

\begin{figure}
(a)\raisebox{-0.9\height}{\includegraphics[width=0.47\textwidth]{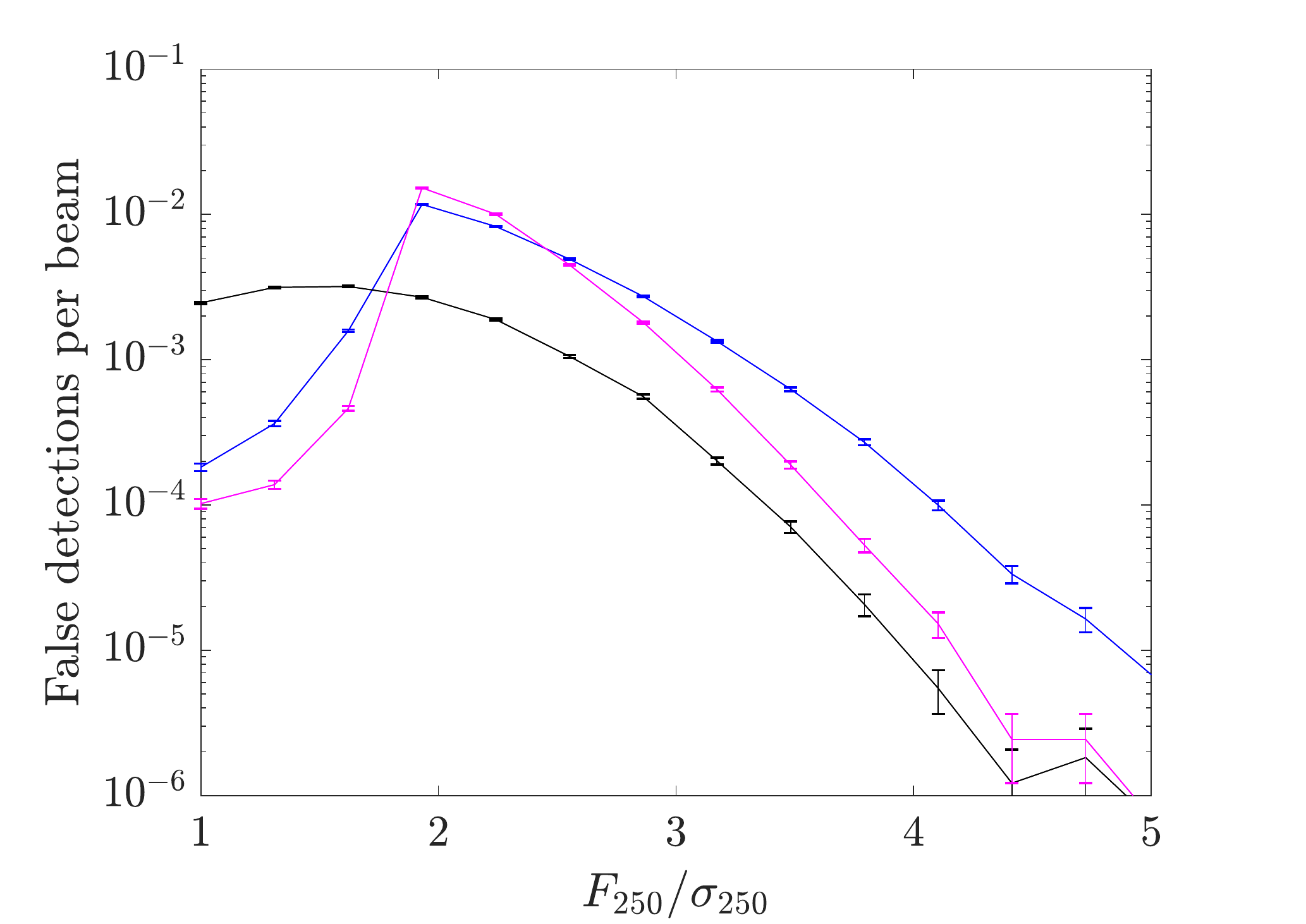}}\\
(b)\raisebox{-0.9\height}{\includegraphics[width=0.47\textwidth]{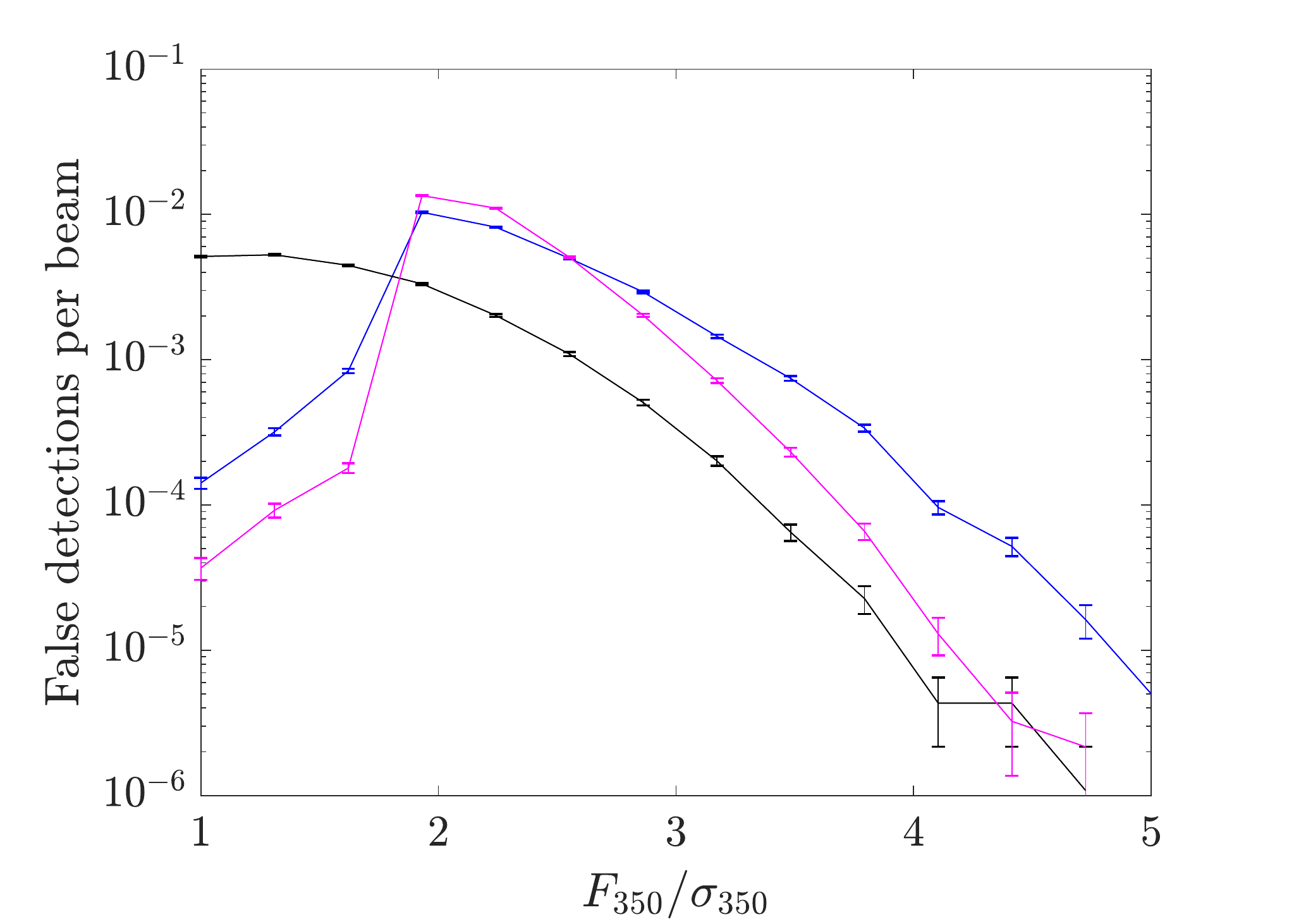}}\\
(c)\raisebox{-0.9\height}{\includegraphics[width=0.47\textwidth]{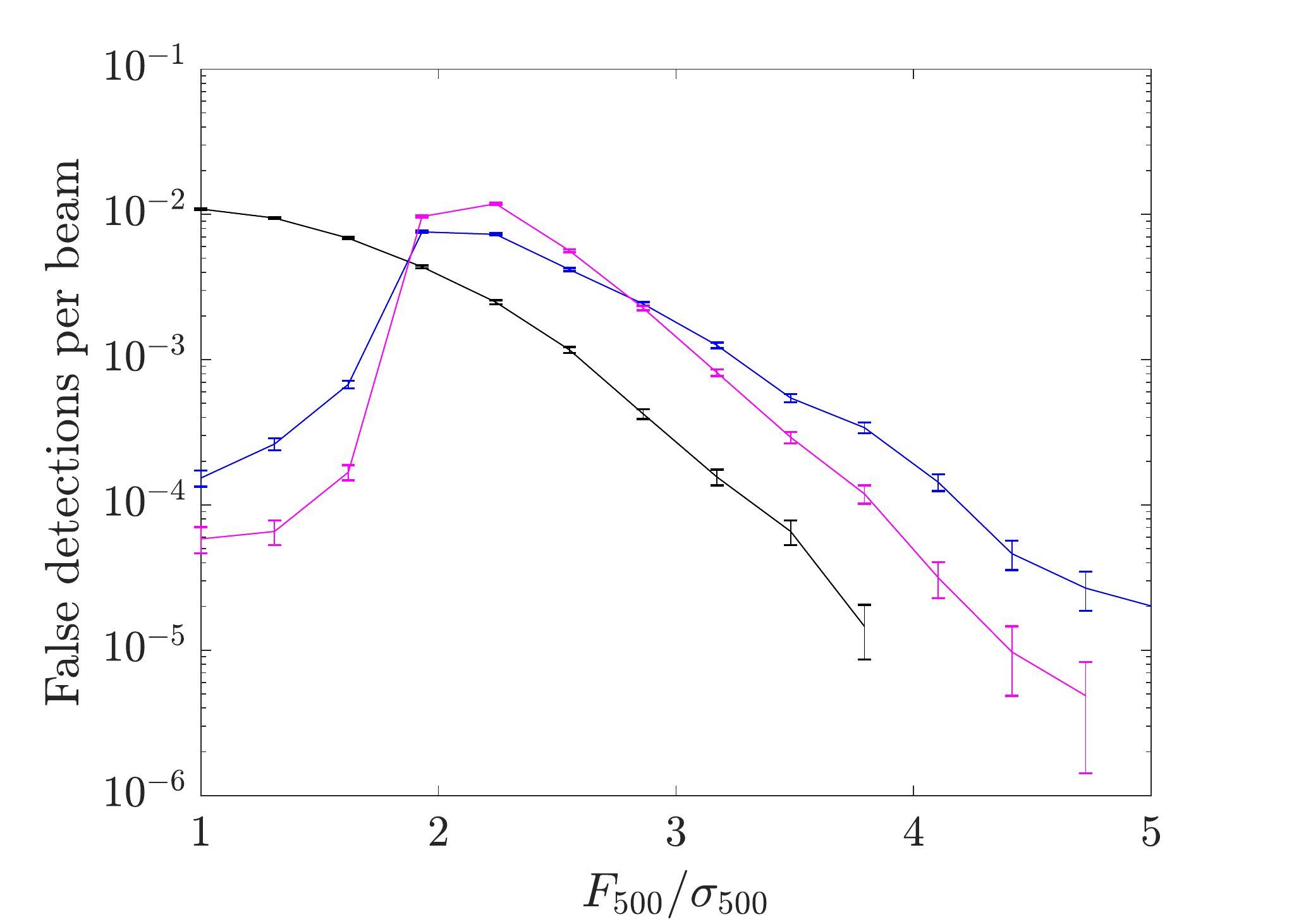}}
\caption{
  The number of false detections per beam as a function of signal to
noise for simulations with confusion noise. The blue lines are for
source detection using the appropriate single-band priors and PSF
filtering; the magenta lines are for source detection using the
single-band priors, but with the matched filters; and the black lines
use the matched filters and the equal weighting of bands. In the
250\mic band, using the matched filter reduces the false detection
rate by a factor 8 at the 4-$\sigma$ limit. Using the flat prior
source detection reduces it by a further factor of 3.  In the 350 and
500\mic bands the gain is roughly a factor 10 between 2 and
4-$\sigma$.
\label{fig:false_conf} }
\end{figure}

As shown in Figure\,\ref{fig:false_conf}, using the matched filter
reduces the number of false detections by a factor 8 at the
4-$\sigma$ limit for the 250\mic band. Using the flat prior as well
reduces the rate by a further factor 2. In the 350\mic and 500\mic
bands the matched filter provides similar reductions in the false
detection rate. 

\begin{figure}
(a)\raisebox{-0.9\height}{\includegraphics[width=0.47\textwidth]{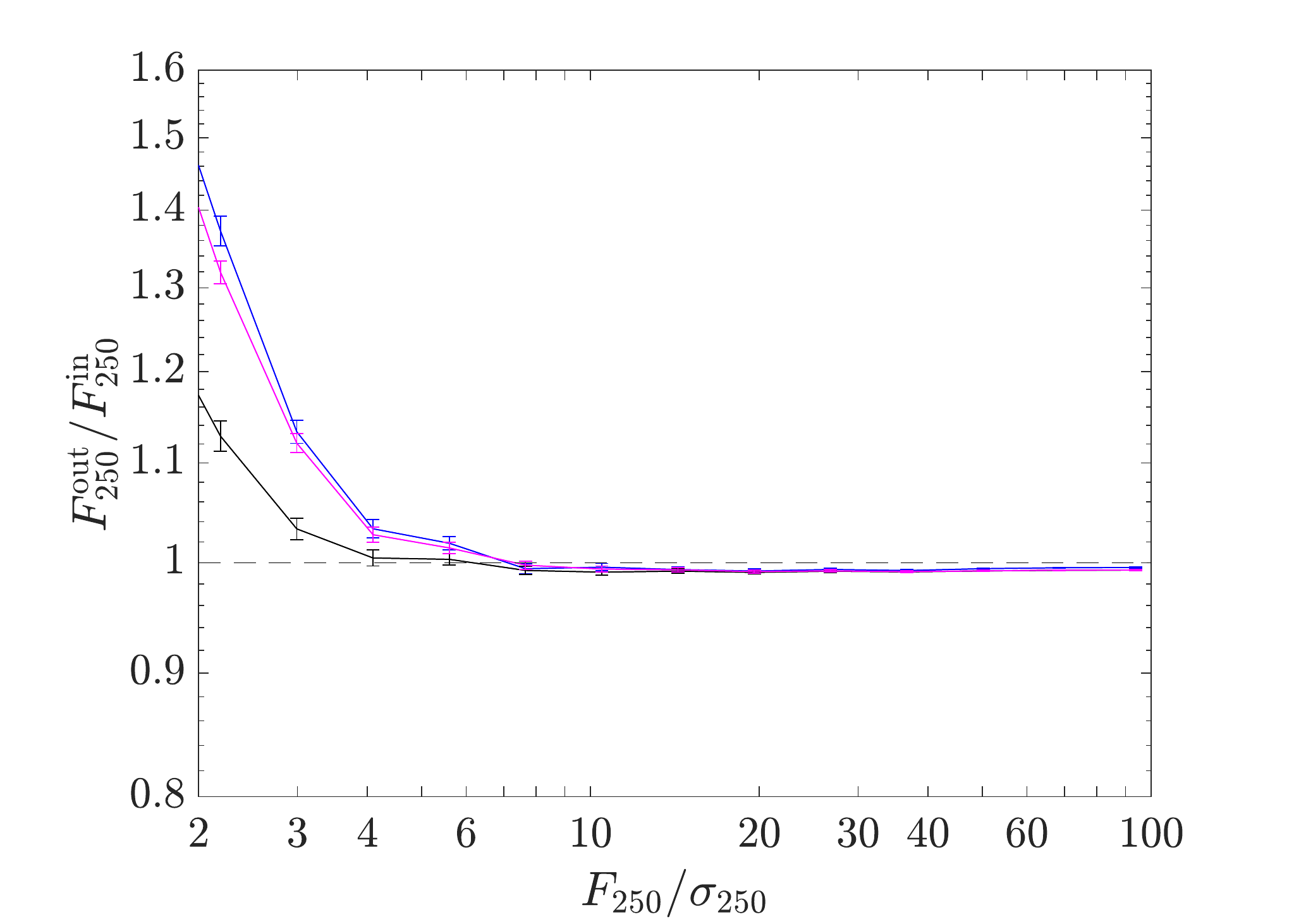}}\\
(b)\raisebox{-0.9\height}{\includegraphics[width=0.47\textwidth]{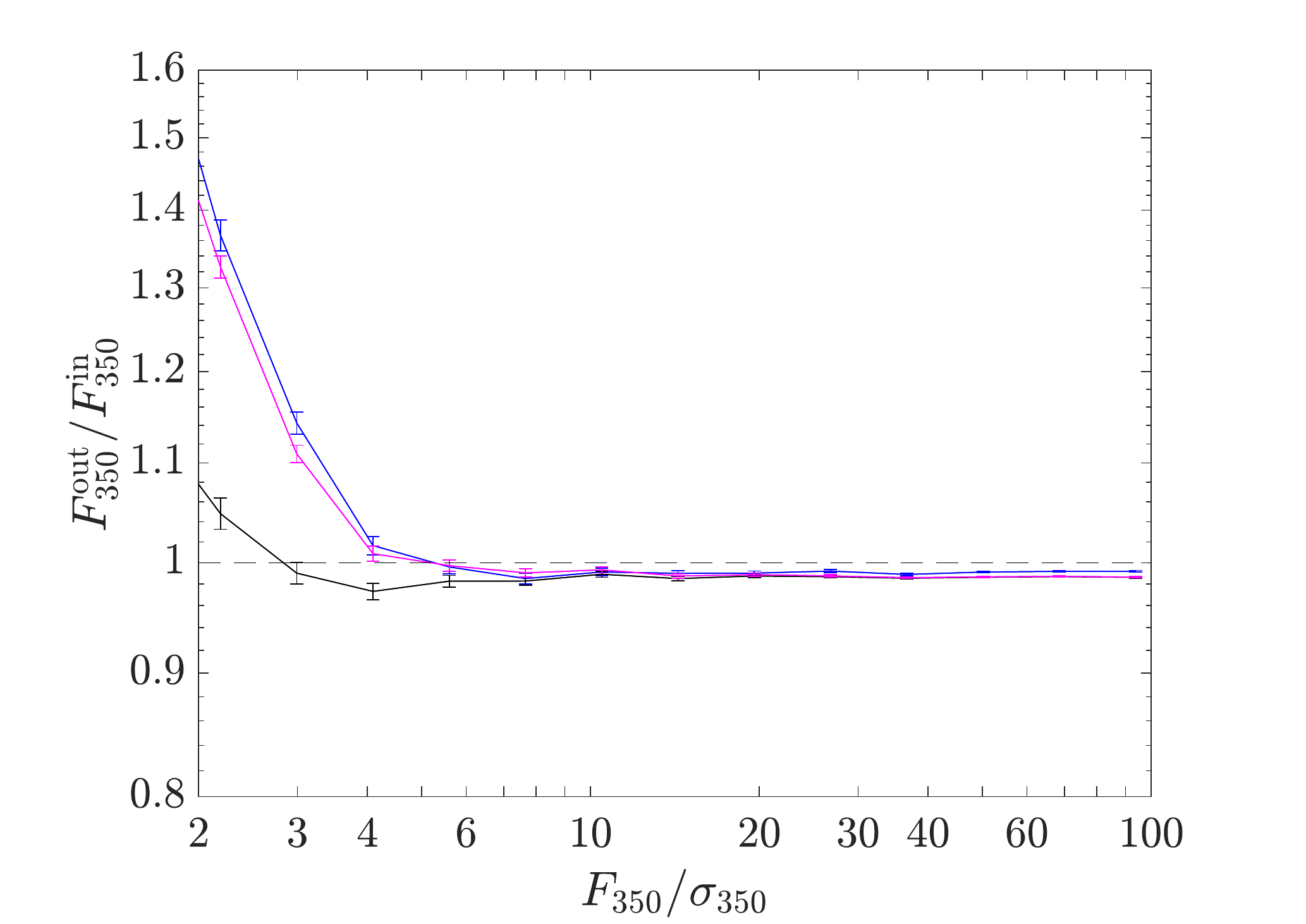}}\\
(c)\raisebox{-0.9\height}{\includegraphics[width=0.47\textwidth]{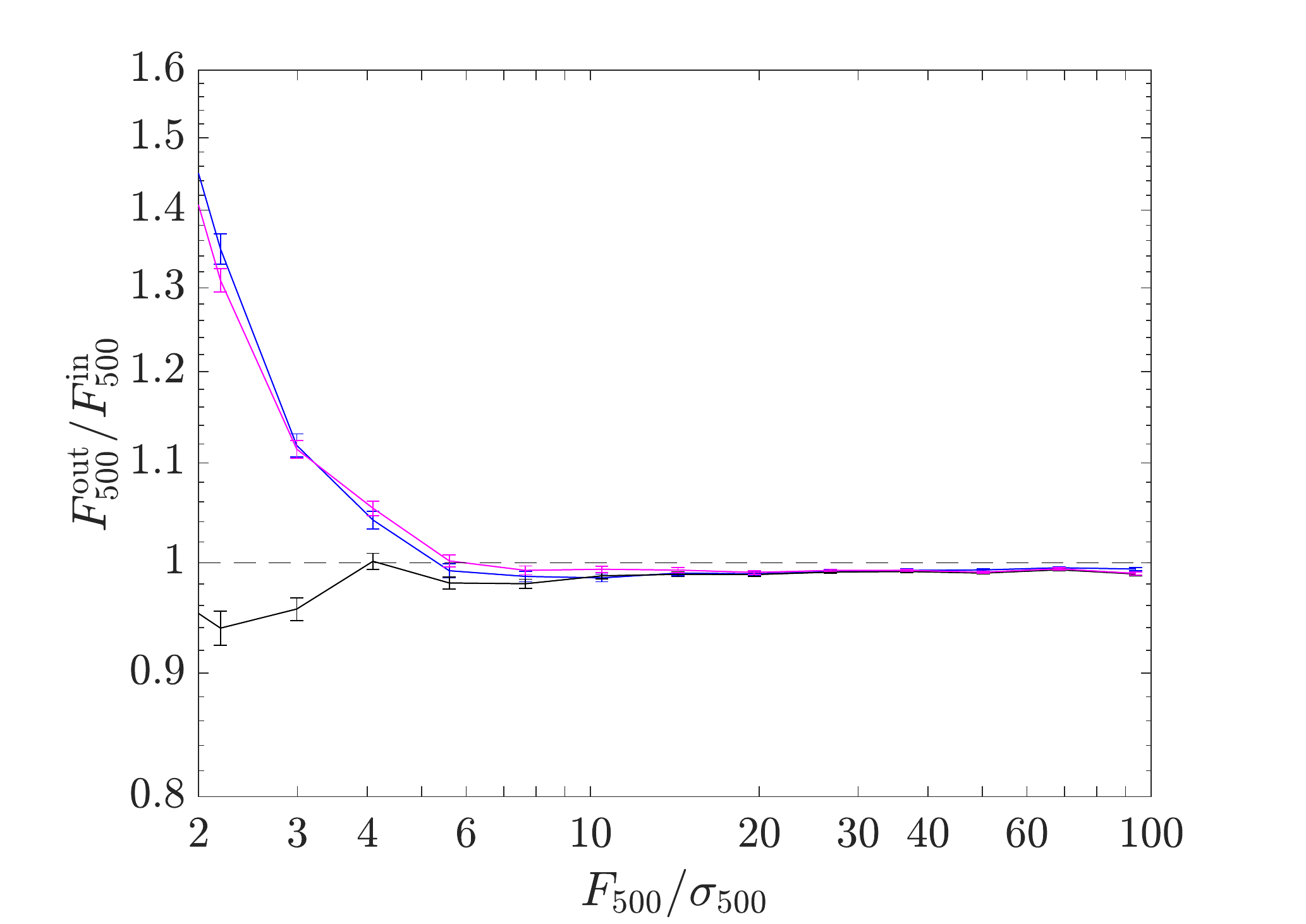}}
\caption{   The ratio of mean measured flux compared  to the mean input flux as
  a function of input signal to noise for different source detection
  priors for simulations with confusion noise included. The blue lines
  show the  single-band priors with a PSF filter; the magenta lines
  show the single-band priors with  a matched filter; the black lines
  show the flat prior  with  a matched filter.
  Panels (a), (b) and (c) shows the results for the 250\mic, 350\mic and
  500\mic fluxes respectively. Using the single-band source detection
  leads to significant flux boosting in the measured fluxes. The flat
  prior reduces the boosting  effect at fainter fluxes.  
  \label{fig:flux_conf} }
\end{figure}

Figure\,\ref{fig:flux_conf} shows the ratio of measured to input fluxc
when confusion noise is included. The behaviour is very similar to
that shown in Figure\,\ref{fig:fluxes} for simple Gaussian noise.  At
high signal to noise there is a small ($\sim$0.5 per\,cent)
underestimate of flux due the peak pixelization issues discussed in
Section~\ref{sec:sub-pixel}.  Near the detection limit, the
single-band detection leads to significant flux boosting.  Using the
flat prior to detect sources reduces this effect. For the 500\mic
band, the positional errors again mean that the fluxes are
underestimated, but the effect is slightly smaller than for Gaussian
noise because the confusion noise has less power on the small scales
that directly affect the position estimates.

\section{Summary}

We have presented a simple approach to detecting sources in data which
consists of multiple broad-band images. For high signal to noise
sources, fitting a Gaussian to estimate the source position and using
bi-cubic interpolation to estimate fluxes significantly improves the
accuracy over single pixels estimates.  Combining multiple bands in an
optimal way before detecting images leads to a significant improvement
in sensitivity to faint sources, a reduction in the number of false
detections, and an improvement in positional accuracy. Using a matched
filter which accounts for confusion noise improves the signal-to-noise
ratio of individual flux measurements and so further improves the
source detection reliability. Together the two modifications provide
catalogues a factor two to three deeper than possible with a standard
single-band PSF filter approach.

\section*{Acknowledgements}

LD and SJM acknowledge support from the European Research Council
(ERC) in the form of Consolidator Grant {\sc CosmicDust}
(Proposal ERC-2014-CoG-647939, PI H\,L\,Gomez), and support from the ERC in the
form of the Advanced Investigator Program, COSMICISM (Proposal ERC-2012-ADG-321302, PI R.J.Ivison).


\label{lastpage}

\end{document}